\documentclass[preprint,12pt]{elsarticle}

\usepackage{lscape}
\usepackage{graphicx}

\usepackage{amssymb}
\usepackage[mathlines]{lineno}
\begin{document}

\begin{frontmatter}

\title{Incomplete cooling down of Saturn's A ring at solar equinox: 
Implication for seasonal thermal inertia and internal structure of ring particles}

\author[ucla,jpl]{Ryuji Morishima\corref{cor1}}
\ead{Ryuji.Morishima@jpl.nasa.gov}

\author[jpl]{Linda Spilker}
\author[jpl]{Shawn Brooks}
\author[seti,jpl]{Estelle Deau}
\author[seti]{Stu Pilorz}

\cortext[cor1]{Corresponding author}

\address[ucla]{University of California Los Angeles, Los Angeles, CA  90095, USA}

\address[jpl]{Jet Propulsion Laboratory, Pasadena, CA 91109, USA}

\address[seti]{SETI Institute ,
Mountain View, CA 94043, USA}

\begin{abstract}
At the solar equinox in August 2009, 
the Composite Infrared Spectrometer (CIRS) onboard Cassini showed the lowest Saturn's ring temperatures ever observed. 
Detailed radiative transfer models show that the observed equinox temperatures of Saturn's A ring are much higher than model predictions 
as long as only the flux from Saturn is taken into account.  In addition, the post-equinox temperatures 
are lower than the pre-equinox temperatures at the same absolute solar elevation angle. 
These facts indicate that the A ring was not completely cooled down at the equinox
and that it is possible to give constraints on the size and seasonal thermal inertia of ring particles
using seasonal temperature variations  around the equinox. 
We develop a simple seasonal model for ring temperatures and 
first assume that the internal density and the thermal inertia of a ring particle are uniform with depth. 
The particle size is estimated to be 1-2 m. 
The seasonal thermal inertia is found to be 30-50 Jm$^{-2}$K$^{-1}$s$^{-1/2}$ in the middle A ring whereas it 
is $\sim$ 10 Jm$^{-2}$K$^{-1}$s$^{-1/2}$ or as low as  the diurnal thermal inertia in the inner and outermost regions of the A ring. 
An additional internal structure model, in which a particle has a high density core surrounded 
by a fluffy regolith mantle, shows that the core radius relative to the particle radius is about 0.9
for the middle A ring and is much less for the inner and outer regions of the A ring.  
This means that the radial variation of the internal density of ring particles exists across the A ring. 
Some mechanisms may be confining dense particles in the middle A ring against viscous diffusion.
Alternatively, the (middle) A ring might have recently formed ($<$ 10$^{8}$ yr) by destruction of an icy satellite, 
so that dense particles have not yet diffused over the A ring and regolith mantles of particles have not grown thick.
Our model results also indicate that  the composition of the core is predominantly water ice, not rock.

\end{abstract}


\end{frontmatter}

%\linenumbers

\section{Introduction}
Saturn's main rings consist of many small particles orbiting around the planet.
The range of particle sizes deduced from radio and stellar occultations is roughly 
1 cm to 10 m (Zebker et al., 1985; French and Nicholson, 2000; Cuzzi et al., 2009). 
The composition of ring particles is mostly crystalline water ice and the mass fraction 
of contaminants (e.g., Tholins, PAHs, or nanohematite) is 10\% at most 
(Epstein et al., 1984; Cuzzi et al., 2009) and probably less than 1 \% (Poulet et al., 2003).
A favorable origin of Saturn's rings with such a high content of water ice is stripping of the 
icy mantle of a Titan-sized satellite (Canup, 2010). 
This hypothesis indicates that Saturn's rings are primordial (as old as Saturn itself)
and the initial mass was a few order of magnitude more massive than the present ring mass.  
As a byproduct of viscous diffusion of Saturn's rings, accretion of 
icy satellites occurs at the outer edge of the rings, followed by orbital expansions of the satellites 
due to the torques induced by the rings and the planet (Charnoz et al., 2010, 2011).

While the ring system as a whole may be primordial, local structures of the rings 
generally change on much shorter timescales. 
Particularly, several observational facts indicates that 
the outermost main ring, the A ring, is dynamically young (see Charnoz et al., 2009 for the review). 
The A ring extends from 122,000 km to 137,000 km from the center of Saturn.
Gravitational torques exerted at resonances in the A ring with inner icy satellites, such as  
 Mimas, Janus, and Epimetheus, cause the A ring to collapse to its inner edge in $\sim
 10^{8}$ years (Goldreich and Tremain, 1982; Dones, 1991), unless angular momentum
 is supplied from some other sources.
 The A ring viscosity measured from density wave patterns at satellite resonances is of the order of 100 cm$^2$ s$^{-1}$ or 
 higher (Esposito et al., 1983; Tiscareno et al., 2007). 
This also gives a radial diffusion timescale of $\sim 10^{8}$ years for the A ring width (see also Salmon et al., 2010). 
The high viscosity of the A ring is likely a result of 
wakes produced by gravitational agglomeration of ring particles. 
Self-gravity wakes are observationally suggested from azimuthal brightness asymmetry 
(e.g., Dones et al., 1993; French et al., 2007; Colwell et al., 2006; Hedman et al., 2007; Ferrari et al., 2009) and 
also predicted from $N$-body simulations (Salo, 1995; Daisaka et al., 2001; Salo et al., 2004; Robins et al., 2010).
The apparently young A ring may or may not be compatible with the primordial origin of the ring system.
To clarify the origin and evolution of the ring system, additional observational constraints are invaluable.

Since Saturn Orbit Insertion of Cassini in 2004 to the present, 
the Cassini spacecraft has been observing Saturn's rings in 
broad wavelengths from ultraviolet to radio. 
The Cassini Composite infrared Spectrometer (CIRS) 
has acquired millions of ring spectra in mid- to far-infrared wavelengths (7 $\mu$m - 1 mm) 
at various observational geometries (e.g., Spilker et al., 2006; Altobelli et al., 2008, 2014; Leyrat et al., 2008b).
The wavelength range of CIRS covers the Planck peak of the thermal emission, 
and the effective temperature and the geometric filling factor of a ring are 
derived by applying a Planck fit to a spectrum (see Spilker et al., 2006 for details). 
The typical error in the derived temperature is much less than 1 K. 

Ring temperatures depend on observational geometric parameters.
Ring temperatures vary seasonally primarily due to 
the change of solar elevation angle from -27$^{\circ}$ to 27$^{\circ}$ and 
secondarily due to the change of heliocentric distance of Saturn
(Froidevaux, 1981; Flandes et al., 2010; Pilorz et al., 2014). 
The face illuminated by the Sun (the lit face) is warmer than the unlit face and the difference between 
the lit and unlit face temperatures increases with ring optical depth (Spilker et al., 2006). 
Ring temperatures also vary diurnally due to eclipse cooling in Saturn's shadow (Leyrat et al., 2008b). 
On top of the eclipse cooling, the temperatures of the A ring show quadruple azimuthal modulations 
due to self-gravity wakes (Leyrat et al., 2008b). 
Ring temperatures increase with decreasing solar phase angle (Altobelli et al., 2007; 2009),
indicating that ring particles are not isotropic emitters. 
 
With each temperature variation, different types of constraints on ring structures and particle properties are obtained. 
For example, the diurnal thermal inertia of ring particles is estimated to be $\sim$10 Jm$^{-2}$K$^{-1}$s$^{-1/2}$
for all main rings, using diurnal temperature variation curves including temperatures in Saturn's shadow 
(Ferrari et al., 2005; Leyrat et al., 2008a; Morishima et al., 2011, 2014). 
This low thermal inertia value indicates that  surface regolith of ring particles is 
very fluffy at least for the thickness comparable to the diurnal thermal skin depth, $\sim$ 1 mm. 
However, fluffiness of deep interiors of ring particles cannot be constrained from diurnal temperature variations.  
Very porous regolith on the particle surface is also indicated from the opposition effect seen in photometric phase curves (Deau, 2015). 

CIRS radial scans measured at various observational geometries give 
a radial profile of the bolometric Bond albedo $A_{\rm B}$  (Morishima et al., 2010).
The value of $A_{\rm B}$ is found to be correlated with the ring optical depth, $\tau$: 
$A_{\rm B} = 0.1-0.4$ for the C ring ($\tau \sim 0.1$),    $A_{\rm B} = 0.5$ for the A ring ($\tau \sim 0.5$),
and $A_{\rm B} = 0.6-0.7$ for the B ring ($\tau \ge 0.7$). 
This correlation probably suggests that Saturn's rings have been continuously polluted 
by meteoroid bombardments and that pollution is more effective for optically thinner rings
(Cuzzi and Estrada, 1998; Elliot and Esposito, 2011).
The radial variation of $A_{\rm B}$ inside the A ring is found to be quite small (Morishima et al., 2010). 
This is consistent with the small radial variation of visual and near-infrared reflectances 
of the A ring (Porco et al., 2005; Hedman et al., 2013).

At the solar equinox in August 2009, Saturn's rings revealed the lowest temperatures ever observed;
there was no temperature difference between the southern and northern faces (Spilker et al., 2013). 
At the equinox, as the Saturn flux is dominant, the ring temperature decreases with increasing saturnocentric distance. 
The equinox temperature is found to be relatively higher at optically thinner regions, because ring particles in thin rings  
can be heated by both the southern and northern hemispheres of Saturn. 
The observed equinox temperatures can be well reproduced by the models used in Spilker et al. (2013), except for the A ring.
They applied a multi-particle-layer model developed by Morishima et al. (2009) to the equinox data of the A ring 
and found that the observed A ring temperatures are much higher than model predictions in which 
only the energy source at the equinox is assumed to be the thermal and solar-reflected fluxes from Saturn.
The equinox temperature anomaly is particularly prominent in the middle A ring;
almost all the the radial scans taken at the equinox show a temperature peak around 129,000 km for the A ring.

In this paper, we examine two possibilities for the equinox temperature anomaly of the A ring.  
The first one is that the multi-particle-layer structure assumed in Morishima et al. (2009) is inappropriate for the A ring.
Since the radial peak location of the equinox temperature coincides with the peak location of the amplitude 
of photometric azimuthal brightness asymmetry caused by self-gravity wakes 
(Dones et al., 1993; French et al., 2007), the discrepancy between the modeled and 
observed equinox temperatures may be due to wakes that are not taken into account in Morishima et al. (2009). 
In Morishima et al. (2014), we have developed a new thermal model in which wakes are represented 
by infinitely long elliptical cylinders, originally introduced by Hedman et al. (2007).  
This model can reproduce the observed azimuthal temperature modulation, 
which is found to be caused by the variation of the geometric filling factor seen from the Sun. 
In the present paper, we apply the wake model of Morishima et al. (2014) to 
the A ring data at the equinox. We also perform some additional parameter studies 
using the multi-particle-layer model of Morishima et al. (2009), 
because the entire parameter space has not been exhausted in Spilker et al. (2013). 

The second possibility, for which we spend much more time than the first one,
is incomplete cooling down of A ring particles at the equinox due to the effect of a seasonal thermal inertia. 
If the thermal inertia of the interior of a ring particle 
is high, the interior cools more slowly than the particle surface. 
The heat transport from the interior  warms up the surface, which leads to 
a higher ring temperature than a model prediction without this seasonal effect.
Therefore, if the equinox temperature anomaly is really due to the seasonal effect,
we are able to give constraints on the seasonal thermal inertia, using seasonal 
temperature variation curves including the equinox data.
The thermal skin depth associated with the seasonal effect is a few order of magnitude 
larger than the diurnal skin depth, and is comparable to the ring particle size.
Thus, combining with the diurnal thermal inertia, it is possible to clarify
how the fluffiness of ring particles changes with their depth.
The seasonal thermal inertia can be constrained only if the particle size 
is larger than the seasonal thermal skin depth. Otherwise, the temperature of the entire particle down to its center 
quickly adjusts to the equilibrium temperature determined from external heat sources. 
Thus, either  the particle size or the seasonal thermal inertia would be constrained.

In Section~2, we explain our methodology. 
In Section~3, we analyze the CIRS equinox data, 
using two sophisticated radiative transfer models which assume extremely 
different ring structures.
We will show that the equinox temperatures for the A ring are much higher than the model estimates 
regardless of ring structure assumed in the models, as long as only the flux from Saturn is taken into account. 
In Section~4, we analyze seasonal temperature variation using all the existing CIRS data.
We introduce a simple seasonal temperature model 
in which the thermal diffusion equation is numerically solved for the temperature evolution of the particle interior. 
We adopt two models for the internal structure of a particle: the first one assumes that
the internal density and the thermal inertia are uniform with depth whereas the second one assumes that a particle has a dense core 
covered by a fluffy regolith mantle. 
The particle size, the seasonal thermal inertia, and the core size estimated from model fits are presented. 
In Section~5, we discuss implication from our results.
The conclusions are given in Section~6.

\section{Methodology}
\begin{figure}
\begin{center}

\includegraphics[width=.9\textwidth]{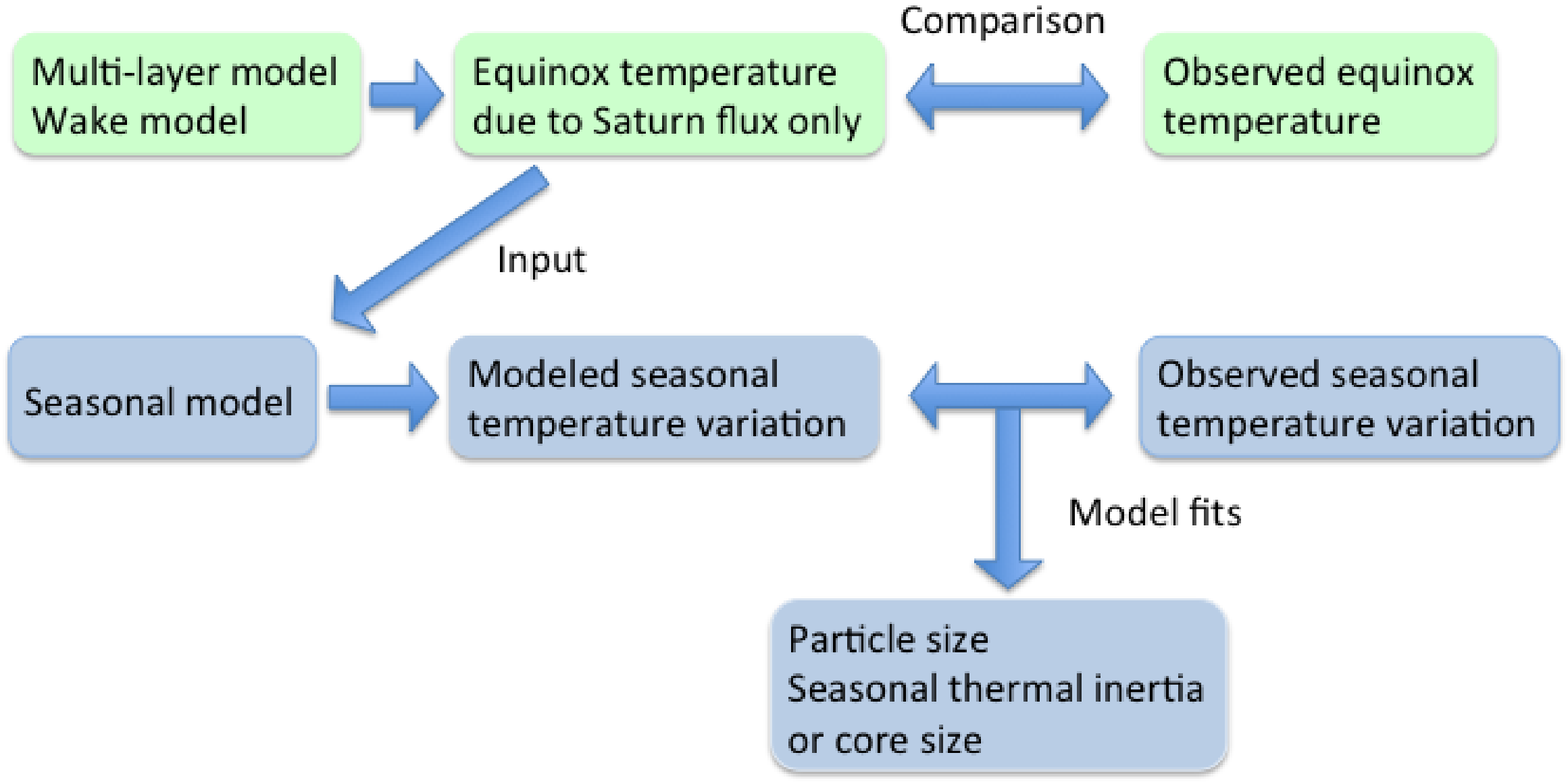}

\end{center}
Fig.~1. Schematic diagram of the methodology. 
\end{figure}

We briefly review our methodology (Fig.~1) to familiarize the reader with the highlights from previous papers 
that are applicable to this work. 
In the present paper, we analyze two different data sets, the equinox data and the seasonal data, using three different thermal models:
the multi-particle-layer model (Morishima et al., 2009), the mono-layer wake model (Morishima et al., 2014), and the seasonal model.

As mentioned in Section~1, the first two models that precisely conserve the energy are used for comparison with the equinox data (Section~3).
Since both models explicitly give the temperature distribution of the particle surface (or the wake surface) by solving 
the radiative heat transfer problem,  they are applicable to any observational geometries. 
These models also take into account diurnal temperature variations by solving the thermal diffusion equation 
for the thin layers near the particle (wake) surface.
However, these models calculate energy balance of the ring at a fixed solar elevation angle, 
neglecting effects due to the seasonal thermal inertia. 
Including  the seasonal effects in these two models is beyond the scope of the present paper,
partly because of difficulties in coding. 
Even if we develop such models,  simulations are most likely computationally very intense because
we need to perform radiative transfer calculations at every solar elevation angle.

% and because an extremely large number of 
%time steps for the thermal diffusion equation is necessary to follow the thermal evolution over Saturn year.

Instead, we introduce a simple seasonal model for analysis of the seasonal data (Section~4). 
The model ignores diurnal variation of the incoming flux but 
takes into account the seasonal effect by solving the thermal diffusion equation for 
the entire particle down to its center. 
The model calculates the spherically averaged temperatures of a ring particle 
without considering a temperature distribution over the particle surface. 
The spherically averaged solar flux onto the particle is given by the analytic formula in Altobelli et al. (2008). 
The formula takes into account mutual shading assuming a mono-particle-layer ring.
In the model, mutual heating is also represented by a simple analytic formula (Froidevaux, 1981; Ferrari et al., 2005).
Use of these analytic formulas allows us to omit any computationally intense radiative transfer calculations.
The problem is that the simple seasonal model itself does not give an accurate equinox temperature with the Saturn flux only, 
while it is essential in the estimation of the seasonal thermal inertia.
Therefore, we use the equinox temperatures derived by the two radiative transfer models (Morishima et al., 2009, 2014). 
Fortunately, as we will show, both radiative transfer models give similar equinox temperatures 
even though very different ring structures are assumed. Thus, the equinox temperature for the mono-particle-layer model
is expected to be similar as well.

Readers familiar with these previous studies may want to skip Section~3 and start reading from Section~4. 

%\section{Incomplete cooling down of the A ring at equinox}
%In this section, we will show evidence of incomplete cooling down of the A ring at the equinox.
%This is suggested by two facts.
%The first one is the observed equinox temperatures much higher than model predictions. 
%The second one is asymmetry in the seasonal temperature variation around the equinox.

\section{Comparison between modeled and observed equinox temperatures}

\subsection{Equinox data}
The solar equinox occurred on August 11th in 2009. During the equinox period (August 11th-13th),
CIRS obtained 15 radial scans (Spilker et al., 2013). The absolute solar elevation angle during the equinox period   
was between 0 and 0.036$^{\circ}$. The very low solar elevation angle guarantees that the direct solar illumination is negligible. 
The spacecraft elevation angles were about 20$^{\circ}$ for all the scans. 

We analyze the equinox data in two forms.   
The first one is mean temperatures observed around 129,000 km ($\pm 500$ km) with various observational geometries. 
The equinox temperature at a certain distance from Saturn increases with decreasing Saturn phase angle (Saturn-ring-observer angle)
indicating that ring particles are not isotropic emitters, although the geometry dependence is very weak for the A ring.  
Instead of the Saturn phase angle, we use the spacecraft hour angle in the rotating frame, 
$L_{\rm S/C,R}$,   introduced in Spilker et al. (2013),  to examine the temperature asymmetry between the leading and  trailing sides of the ring.
The angle is give as 
$L_{\rm S/C,R} = L_{\rm S/C} - L_{\rm R}$,  
where $L_{\rm S/C}$ is spacecraft hour angle around a footprint and $L_{\rm R}$ is the ring local hour angle around Saturn.
In this frame, the direction of Saturn seen from the ring is always at  $L_{\rm S/C,R} = 12$ h $(= 180^{\circ})$, 
and the leading and trailing sides of the ring are seen when $L_{\rm S/C,R} < 12$ h and $L_{\rm S/C,R} > 12$ h, respectively.
Since the spacecraft elevation angles are almost the same for all the equinox scans, 
the Saturn phase angle simply increases as $L_{\rm S/C,R} $ recedes away from 12 h.
The second one is the temperatures at various saturnocentric distances. 
We choose two radial scans that give the highest and lowest 
mean temperatures at the equinox as representatives. Radial temperature profiles of other scans 
are very similar to these two scans.
More details for the equinox data set are found in Spilker et al. (2013).   

\subsection{Radiative transfer models}

We test whether thermal models can reproduce the equinox temperatures.
Two radiative transfer models are employed: a multi-particle-layer model (Morishima et al., 2009) and 
a mono-layer wake model  (Morishima et al., 2014). 
In the models, Saturn's thermal emission and solar illumination reflected by Saturn are taken into account, 
and direct solar illumination is ignored.
The effective surface temperature of Saturn is 96.67 K (Li et al., 2010) and Saturn's oblateness is 0.1.

\subsubsection{Multi-particle-layer model}

In the multi-particle-layer model of Morishima (2009), 
the vertical thickness of a ring is assumed to be much larger than the particle size.
The model numerically solves the classical radiative transfer equation (Chandrasekhar, 1960; Modest, 2003)
using the multi-stream method (we use 60 streams). 
We adopt the plane-parallel approximation so rings are horizontally homogeneous. 
The model takes into account heat transport due to particle motion in the vertical and
azimuth directions by following temporal thermal evolution of thousands of ring particles.
 It assumes, instead of an actual continuous size
distribution, a bimodal size distribution consisting of fast and slow rotators.
Fast rotators are small-rapidly
spinning particles having spherically symmetric thermal structure whereas
slow rotators are  large non-spinning Lambertian particles.
The ring effective temperature is derived by summing up the thermal emission of all ring particles toward the observer, 
taking into account the attenuation factor due to other particles on the line of sight. 

In the present work, we consider two extreme 
cases of particle spins: fast rotators only and slow rotators only.
In the model, we adopt the following parameters:  the infrared emissivity $\epsilon_{\rm IR} = 1.0$, 
the thermal inertia $\Gamma = 10$ Jm$^{-2}$K$^{-1}$s$^{-1/2}$, and the bolometric Bond albedo 
$A_{\rm B} = 0.5$. The latter two parameters are estimated by Morishima et al. (2010, 2011).

\subsubsection{Mono-layer wake model}

In the mono-layer wake model of Morishima et al. (2014), the temperature distribution 
over the wake surface is calculated instead of considering the temperatures of individual particles.
Wakes are approximately represented by infinite, parallel, regularly spaced, opaque 
elliptical cylinders, with vertical thickness $H$, horizontal width $W$, and wavelength $\lambda$, as adopted in Hedman et al. (2007). 
Our model ignores interwake particles for simplicity. 
The surface of an elliptical cylinder is divided into multiple facets and 
interactions between facets (mutual heating and multiple scattering of photometric light) 
are calculated using a radiosity method.
As discussed in Morishima et al. (2014), it is likely that some heat transport occurs
within wakes; this may be due to vertical motion of particles,
or particle spins may play a role if wakes are represented by a
mono-particle-layer.
To represent heat transport approximately, 
the incoming flux is smoothed over the wake surface and its degree $f_{\rm mix}$ ($0 \le f_{\rm mix} \le 1$)
is given as a parameter. If  $f_{\rm mix} = 1$, the wake surface is isothermal. 
The ring effective temperature is calculated by summing up the thermal emission of facets seen from the observer.

We use  $f_{\rm mix} = 0.8$, the wake bolometric reflectance $A_{\rm V} = 0.35$,
and $\Gamma = 10$ Jm$^{-2}$K$^{-1}$s$^{-1/2}$ as estimated by Morishima et al. (2014) unless 
otherwise noted ($f_{\rm mix}$ is varied in Fig.~4). The pitch angle of wakes is assumed to be 30$^{\circ}$.
A similar value is derived by Ferrari et al. (2009). 

\subsection{Equinox temperature results}

\subsubsection{Munlti-particle-layer model results}

\begin{figure}
\begin{center}
\includegraphics[width=.65\textwidth]{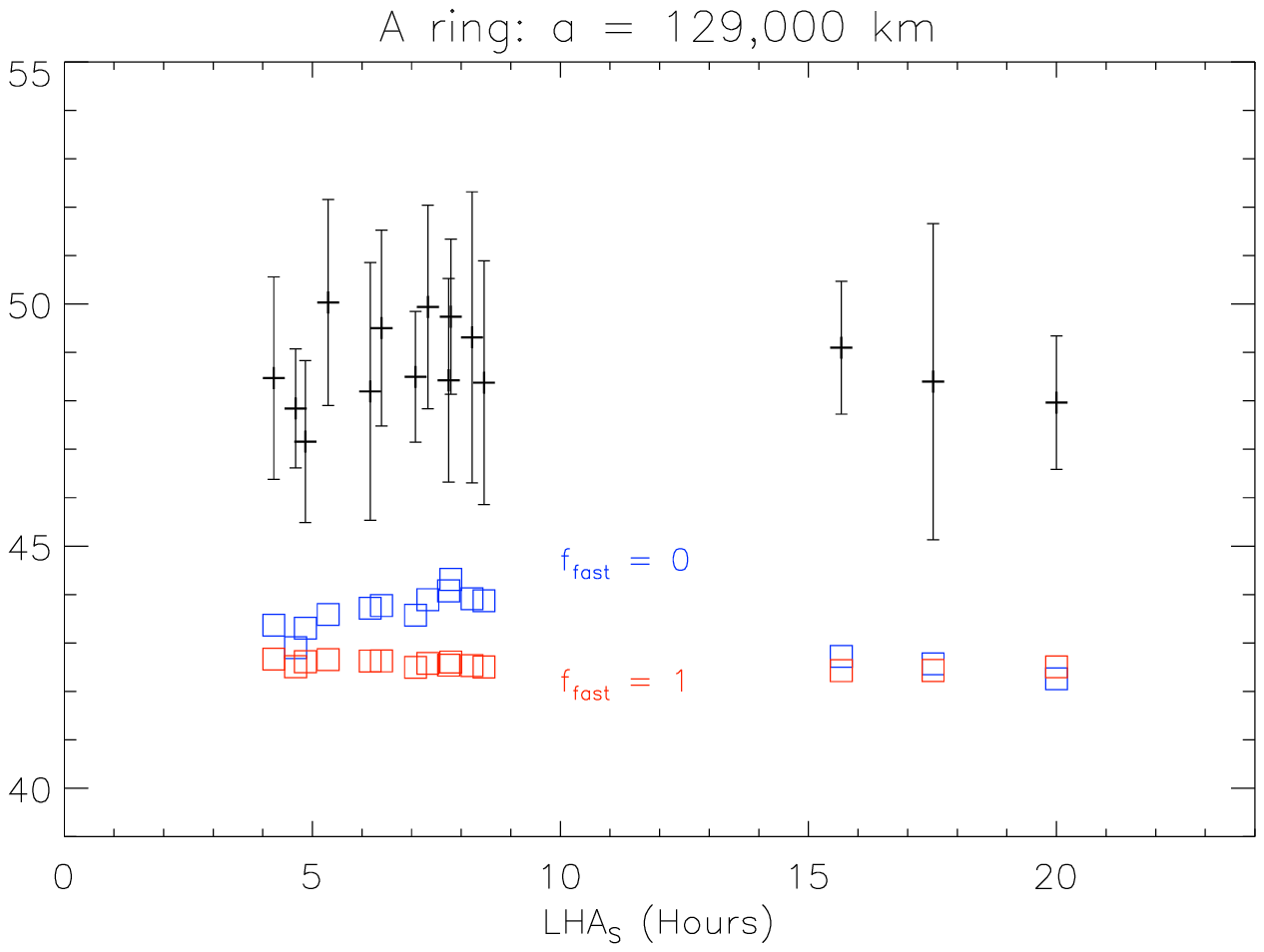}
\end{center}

Fig.~2. Equinox temperature variation with local hour angle of Cassini in the rotating frame.
The saturnocentric distance of the ring is $129,000$ km and the radially averaged ($\pm 500$ km) temperatures are shown.
The direction of Saturn is 12 hr in this frame. 
Black pluses (with error bars) are the observed temperatures and blue (fast rotators) and red (slow rotators) 
squares are temperatures from the multi-particle-layer model.

\end{figure}

\begin{figure}
\begin{center}

\includegraphics[width=.65\textwidth]{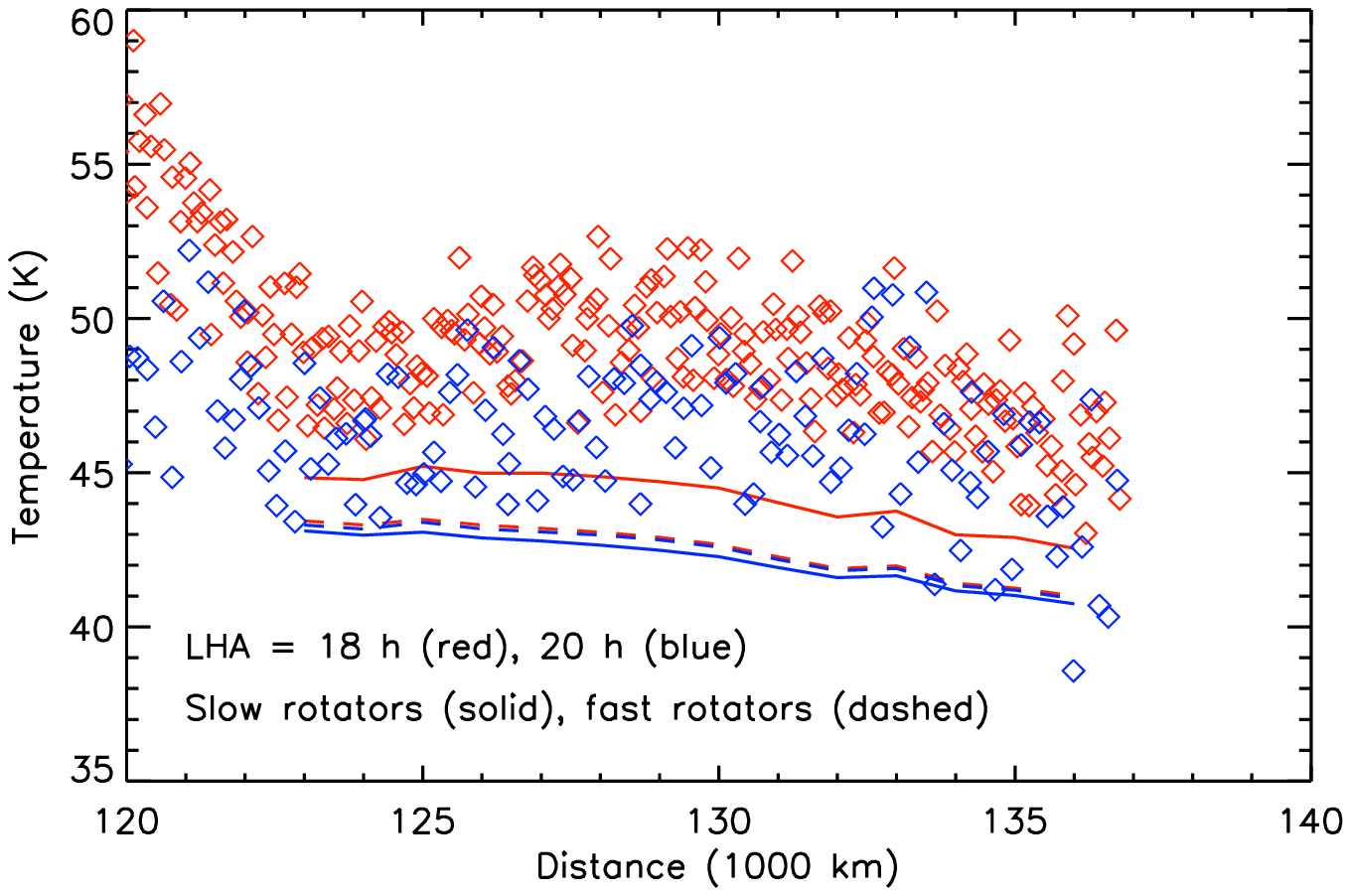}

\end{center}
Fig.~3. Radial profiles of equinox temperatures. 
Individual diamonds are temperatures retrieved by individual spectra.
Two scans that give the highest (red diamonds) and lowest (blue diamonds)
mean temperatures are shown.
Radial temperature profiles calculated by the multi-particle-layer model are also plotted by solid (slow rotators) and dashed (fast rotators) lines.

\end{figure}

Figure~2 shows the variation of the equinox temperatures with $L_{\rm S/C,R}$ at $a = 129,000$ km. 
The observed temperatures in the figure are the same with those shown in Spilker et al. (2013) (their Figs.~3 and 7).  
In the model, we adopt the ring optical depth $\tau$ of  0.5.
Both cases of particle spins give equinox temperatures that are much lower than the observed temperatures 
although slow rotators give slightly higher temperatures than fast rotators.
Similar results were previously presented in Spilker et al. (2013) in which 
the total cross sections of fast and slow rotators are assumed to be roughly the same, 
as suggested by Morishima et al. (2010, 2011).

Figure~3 shows the radial variation of the equinox temperatures at the two different geometries.
In the model, we use the optical depths from Voyager PPS (Esposito et al., 1983) smoothed 
over radial bins with a 1,000 km width, avoiding the Encke division.  
The observed radial profiles show a temperature peak around 129,000 km whereas the model 
predicts that the temperature gradually decreases with increasing distance. 
The ring normal optical depth is known to depend on observational geometry 
due to wakes (Colwell et al., 2006, 2010; Hedman et al., 2007).
If we use larger optical depths given by some of Cassini occultations,
slightly lower temperatures are obtained due to an enhanced mutual shading effect
while the relative radial temperature variation is almost unchanged.

Overall, the model cannot reproduce not only the absolute temperature but also 
the radial profile of the equinox temperatures. 
If we adopt an infrared emissivity lower than
unity, that can probably give slightly higher temperatures. Since the current multi-particle-layer code 
does not include scattered thermal light, we cannot directly check cases with $\epsilon_{\rm IR} < 1.0$.
On the other hand, we can roughly infer its effect from the wake model which includes scattered thermal light. 
We see that the equinox temperatures for $\epsilon_{\rm IR} = 0.9$ are higher 
than those for $\epsilon_{\rm IR} = 1.0$ by 1-3 K (Fig.~4). This difference is still insufficient to explain 
the observed temperature anomaly, particularly in the middle A ring. 

\subsubsection{Mono-layer wake model results}

\begin{figure}
\begin{center}

\includegraphics[width=.7\textwidth]{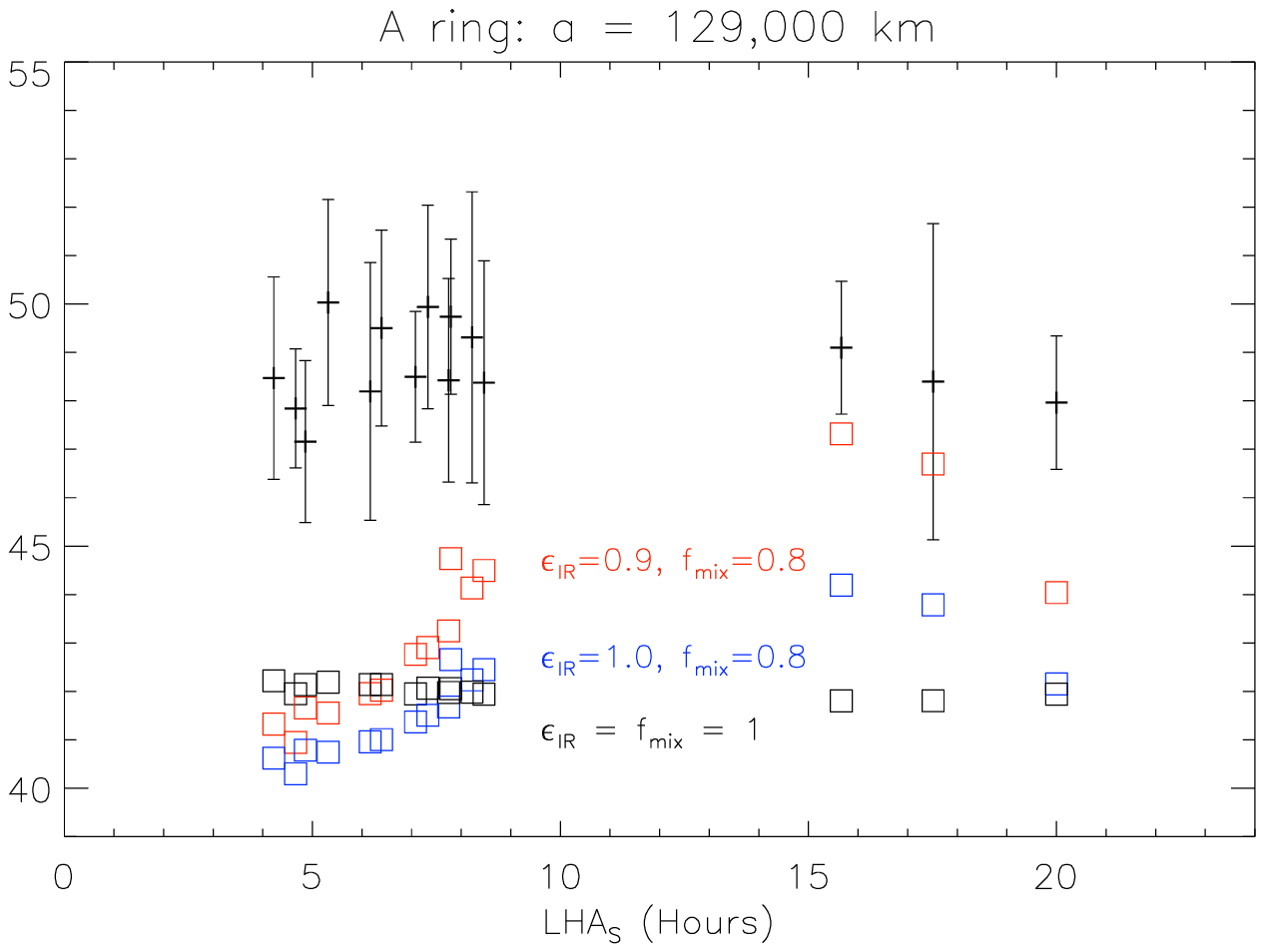}

\end{center}

Fig.~4. 
Same as Fig.~2, but temperatures from the mono-layer wake model are compared with 
the observed temperatures. Dependences on the thermal emissivity $\epsilon_{\rm IR}$  
and the mixing efficiency $f_{\rm mix}$ are examined. 
The height-to-width ratio, $H/W$, is 0.4.

\end{figure}

\begin{figure}
\begin{center}

\includegraphics[width=.8\textwidth]{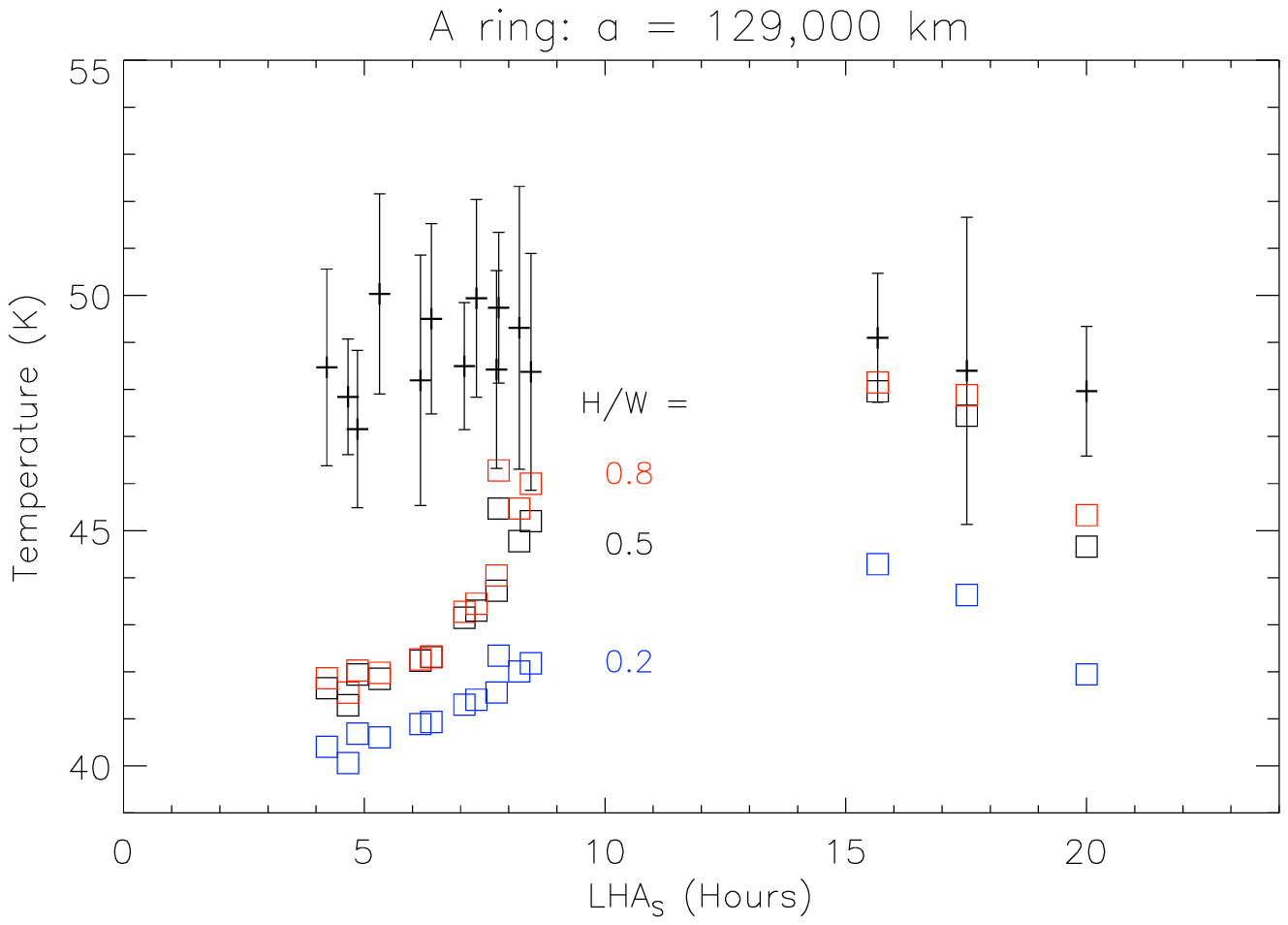}

\end{center}

Fig.~5. 
Same as Fig.~4, but the dependence on height-to-width ratio $H/W$ is examined in the model.
We adopt $\epsilon_{\rm IR} = 0.9$ and $f_{\rm mix} = 0.8$.

\end{figure}

\begin{figure}
\begin{center}

\includegraphics[width=.7\textwidth]{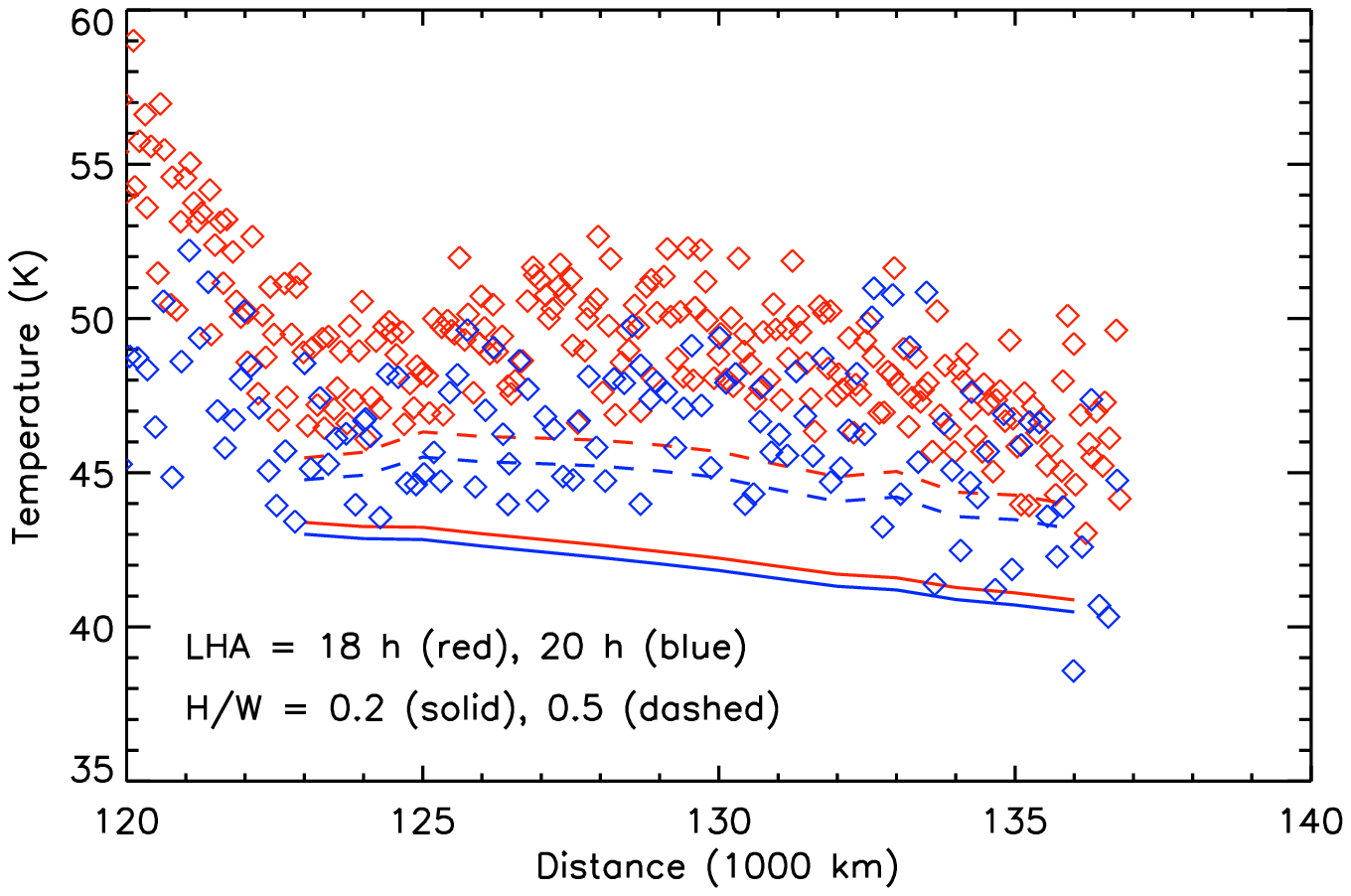}

\end{center}

Fig.~6. 
Same as Fig.~3, but temperatures from the mono-layer wake model are compared with 
the observed temperatures. The results for  $H/W = $ 0.2 (solid) and 0.5 (dahsed) are shown.
We adopt $\epsilon_{\rm IR} = 0.9$ and $f_{\rm mix} = 0.8$.

\end{figure}

Figure~4 shows results of the wake model on top of the observed data similar to what is shown in Fig.~2.
In the model, we adopt $W/\lambda$ = 0.4 and $H/W = 0.4$.
The wake width gives the normal ring optical depth of $\tau = -\log(1-W/\lambda) = 0.51$ which is close to that for the A ring
and $H/W = 0.4$ is the value suggested by Ferrari et al. (2009).
In Fig.~4, we examine dependence on $\epsilon_{\rm IR}$ and  $f_{\rm mix}$.
If $\epsilon_{\rm IR}$ = $f_{\rm mix}$ = 1,
the wake temperature  becomes independent of observational geometry at a certain local hour angle.
For $f_{\rm mix} <1$, the observed ring temperature increases towards $L_{\rm S/C,R} = 12$ h, 
because the surface portions facing towards Saturn have relatively higher temperatures.
However, the temperature averaged over $L_{\rm S/C,R}$ is insensitive to $f_{\rm mix}$.
With decreasing $\epsilon_{\rm IR}$,  the ring physical temperature increases 
due to inefficient thermal emission and a part of infrared emission from Saturn is scattered by wakes toward the observer.
In addition, a low $\epsilon_{\rm IR}$ causes a strong observational-geometry dependence of the temperature 
because of  scattered thermal light.
We find that to reconcile with the observed geometry dependence,  $\epsilon_{\rm IR}$ needs to higher than $\sim$ 0.9.  
The high emissivity is also indicated by spectroscopy (Morishima et al., 2012), 
and is due to strong absorption of water ice at far-infrared wavelengths.
As long as $\epsilon_{\rm IR} > 0.9$, the wake model gives 
equinox temperatures that are much lower than the observed temperatures.

Although $H/W \simeq 0.4$ 
is favored for the elliptical cylinder wake model and CIRS data in the middle A ring (Ferrari et al., 2009; Morishima et al., 2014), 
estimated values of $H/W$ vary with different observations and wake shape models 
(e.g., Cowell et al., 2006; Dunn et al., 2007; French et al., 2007; Hedman et al., 2007).
Thus, we check how the equinox temperature varies with $H/W$ (Fig.~5).
It is found that the equinox temperature increases with $H/W$ as long as $H/W < 0.5$.
For a larger $H/W$, little increase is seen. 
The effective wake cross section seen from Saturn increases with increasing $H/W$, 
as long as other nearby wakes do not hinder illumination from Saturn. That is the case for $H/W < 0.5$.
For $H/W > 0.5$, mutual shading starts to be effective. Similar results are found for cases in which solar illumination 
is the dominant heat source (Morishima et al., 2014). 

Figure~6 shows radial profiles of the equinox temperature derived from the wake models 
for $H/W =0.2$ and 0.5 plotted on top of the observed data similar to what is shown in Fig.~3. 
Again, the optical depth profile from Voyager PPS is used.
It is found that the mono-layer wake model also gives much flatter radial profiles of the equinox temperatures than the observed ones.
Colwell et al. (2006) showed that $H/W$ increases from $\sim$ 0.2 at the inner A ring 
to $\sim$ 0.5 at the outer A ring. If we adopt such a radial variation of $H/W$,
the equinox temperature becomes almost constant over the A ring, inconsistent with the observations.

We conclude that neither the wake model nor the multi-particle-layer model can reproduce
the observed high equinox temperatures and the radial temperature profiles of the A ring
with a plausible range of parameters. 

\section{Particle properties inferred from seasonal temperature variations}
In Section~3, we showed evidence of incomplete cooling down of Saturn's A ring at the equinox.  
This fact allows us to give constraints on  
particle properties such as the size and the seasonal thermal inertia.
In this section, we develop a simple time-dependent model for seasonal ring temperature variations
and compare model results with observations.
   
\subsection{Data selection}
As of the middle 2014, we have about 2,000 CIRS scans. Each scan usually contains thousands of spectra (and temperatures).
We make data selections from the entire set of scans as follows.
We choose data with moderately good spatial resolutions to examine radial variation of particle properties. Specifically, 
data with a radial length of a footprint projected on the ring plane of less than 5,000 km are selected.
While our seasonal model does not take into account observational geometry dependence, the data show variations of temperature with geometry,
particularly when $|B'|$ is high. 
If $|B'| < 2^{\circ}$, we do not remove any data as the geometry and diurnal dependences of temperatures are small enough. 
If $|B'| > 2^{\circ}$, we remove data at solar phase angles lower than $50^{\circ}$ as the low phase data 
show high temperatures. It is more appropriate to calculate the mean temperature by integrating the thermal flux over the solid angle
as done in Pilorz et al. (2014) for the B ring.
Nevertheless, since the A ring shows relatively weak phase dependences (Altobelli et al., 2009),
this simple removal of low phase data has only a minor effect. 
For $|B'| > 2^{\circ}$, we also remove data near Saturn's shadow as the temperatures in the shadow 
are much lower than those around the noon. We use data only with $100^{\circ}< L_{\rm R} <300^{\circ}$.

The temperatures in a single scan (usually a radial scan) are averaged over 14 radial bins with a width of 1,000 km and 
the bin centers from 123,000 km to 136,000 km. The number of data points in a radial bin for a single scan is generally about 10. 
The averaging is done after removing the data outside $3\sigma$, where $\sigma$ is the standard deviation of the temperature. 
If $\sigma$ after removing noisy data is still larger than 3 K, we do not use the scan for this radial bin.

We make further data cleaning. 
Using the model introduced in Section~4.3, we derive the best-fit curves for seasonal temperature variations. 
If the difference between the best fit temperature and the mean observed temperature is larger than 5 K,  we remove
this scan.  These abnormal data most likely suffer from some sort of instrumental errors. 
Since these data are rather rare, model fits are not altered regardless of them.

One may find that we have a great deal of CIRS data before the equinox whereas much less data are available after the equinox 
(e.g., Fig.~7). After the equinox, Cassini changed its orbital inclination relative to the Saturn's equatorial plane from 
moderately high values to nearly zero to observe saturnian satellites. 
Only a small number of ring observations were made soon after the equinox, as 
observational geometries were not suitable for the main rings. The lack of post-equinox data unfortunately 
causes large uncertainties in parameters estimated in the present work.

\subsection{Asymmetry in seasonal temperature curves around equinox}
\begin{figure}
\begin{center}
\includegraphics[width=.6\textwidth]{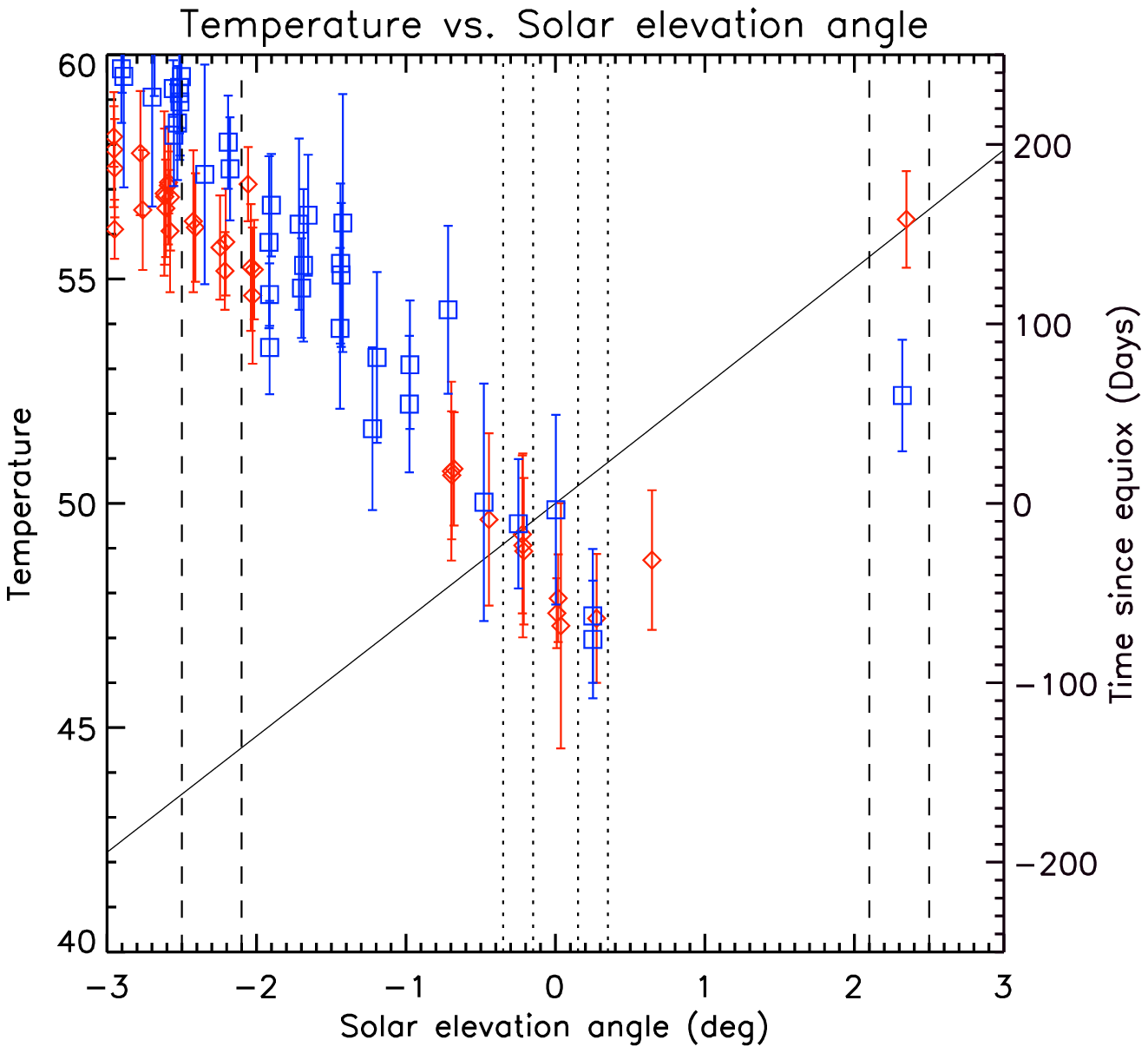}
\end{center}

Fig.~7. 
 Seasonal temperature variation around the solar equinox ($B'$ = 0) at $129,000$ km.
The blue squares and red diamonds are the southern and northern face temperatures, respectively.
The solid line is the time since the equinox.
The temperatures in between dashed and dotted vertical lines are used for measurements of the asymmetry in Fig.~8.

\end{figure}

\begin{figure}
\begin{center}
\includegraphics[width=.5\textwidth]{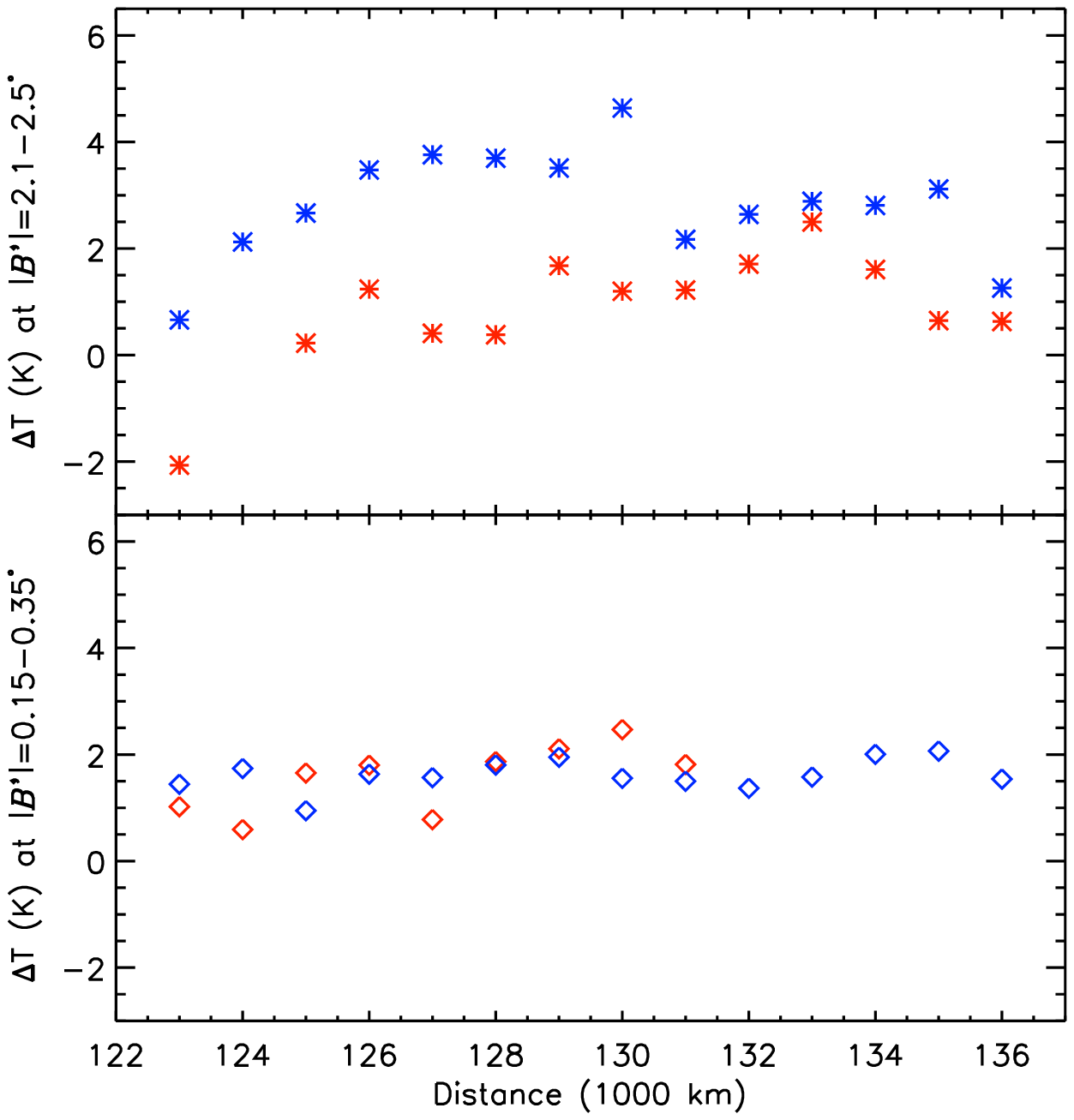}
\end{center}
 
Fig.~8. 
Temperature asymmetry around the equinox.  The upper and lower panels show the temperatures 
before the equinox subtracted by those after the equinox in the ranges of $|B'| = 2.1$-$2.5^{\circ}$
and 0.15-0.35$^{\circ}$, respectively.  The blue and red marks are for the lit and unlit faces, respectively.

\end{figure}

Before moving to modeling, we will show additional evidence of 
incomplete cooling down at the equinox, using seasonal data only. 
Incomplete cooling down means that 
it takes finite time for ring particles to cool down due to a finite thermal inertia and a finite size.
This should also mean that it takes finite time for particles to be warmed up by direct solar illumination 
after the equinox. As a result, it is expected that the ring temperature before the equinox is higher than 
that after the equinox for the same absolute solar elevation angle. This asymmetry 
in seasonal temperature curves is indeed identified. 

Figure~7 shows the seasonal temperature variations of the middle A ring (at 129,000 km) over $\sim$ a year 
around the equinox. The blue symbols are temperatures of the southern face of the ring 
the red symbols are temperatures of the northern face. 
We call a face directly illuminated by the Sun the lit face and the other face the unlit face.
The southern face changes from the lit face to the unlit face at the equinox. 
Asymmetries for both the lit and unlit faces are examined (e.g., comparison between the pre-equinox 
southern face and the post-equinox northern face). 
Figure~7 clearly shows that the pre-equinox temperatures 
are larger than the post-equinox temperatures at the same absolute solar elevation angle
both for the lit and unlit faces.

To quantify asymmetry of temperatures, we take averages of temperatures 
in the intervals between $|B'| = 2.1^{\circ}$ and 2.5$^{\circ}$ and between $|B'| = 0.15^{\circ}$ and 0.35$^{\circ}$, 
and subtract the mean temperature after the equinox from the mean temperature before the equinox 
for the same absolute solar elevation angle.  
Figure~8 shows the temperature differences at various saturnocentric distances for the intervals  
between $|B'| = 2.1^{\circ}$ and 2.5$^{\circ}$ (the upper panel) and between $|B'| = 0.15^{\circ}$ and 0.35$^{\circ}$ (the lower panel).
The blue symbols are for the unlit face and the red symbols are for the lit face.
It is found that the asymmetry is particularly large at the middle A ring, where 
the observed equinox temperatures are much larger than the modeled temperatures. 
The large temperature asymmetry, in addition to the large anomaly in the equinox temperature (Section~3),
strongly indicates the efficiency of cooling down is low in the middle A ring. 
On the other hand, the asymmetry is relatively small at the inner and outer parts of the A ring, where
relatively small anomalies in the equinox temperatures are seen.  
Thus, it is likely that some particles properties vary radially across the A ring. 
Those will be addressed in the following sections.

\subsection{Seasonal model}
\subsubsection{Energy balance for ring particles}

We introduce a seasonal temperature variation model.
The model takes into account time-dependence by solving 
the thermal diffusion equation of a particle.
In the model,  the ring temperature is represented by the surface temperature of a particle and 
the ring particle temperature is assumed to be spherically symmetric.
Therefore, dependency of  ring temperature on observational geometry is not considered although 
the southern and northern face temperatures are calculated separately by giving different solar fluxes to these faces.
The diurnal variation of the incoming flux is also ignored.  
We adopt the analytic formula of the spherically averaged direct solar flux given by Altobelli et al. (2008).  
This formula takes into account mutual shading of particles and accurately reproduces
the flux onto a mono-particle-layer ring  (Froidevaux, 1981).
The model of Froidevaux (1981) is known to accurately reproduce the temperature variation with solar 
elevation for all rings at any observational geometries with an appropriate choice of 
the geometry-dependent particle albedo (Flandes et al., 2010). 
The model requires an accurate Saturn flux that determines the equilibrium equinox temperature that is obtained 
when the time-dependence is ignored. 
For an accurate Saturn flux, we use the equinox temperatures 
given by our own radiative transfer models, instead of the approximate formula given by Froidevaux (1981). 
We use the equinox temperatures averaged over the four radial profiles shown in Fig.~3.   
These temperatures are calculated by the multi-particle-layer model but the mono-layer wake model gives similar temperatures (Fig.~6).

To develop a model of the seasonal temperature variation of a ring,
we first need to determine the heliocentric distance $r_{\rm S}$ and the solar elevation angle $B'$ as a function of time, $t$.
We simply assume that Saturn has a Keplerian orbit around the Sun. The details are given in Appendix~A.

Time evolution of the particle temperature $T$ is  solved by the one-dimensional thermal diffusion equation: 
\begin{equation}
\frac{\partial T}{\partial t} = \frac{1}{\rho C}\frac{1}{r^2} 
\frac{\partial}{\partial r}
\left(r^2 K \frac{\partial T}{\partial r}\right), \label{eq:difeq}
\end{equation}
where $K$ is the thermal conductivity, $\rho$ is the internal density, and $C$ is the specific heat.
We use a fixed value, $C = 450$ J kg$^{-1}$ K$^{-1}$, which is the value of water ice at $\sim 50$ K (Shulman, 2004),
whereas $K$ and $\rho$ are varied in the model.
We ignore the temperature dependence of $C$ and $K$ for simplicity. 
This is probably a safe assumption as the time-dependence is important only around the equinox and 
the temperature variation during that period is not large.
At high solar elevation, the ring temperature is very close to the equilibrium temperature given with $K = 0$.

The boundary condition at the surface  is given by 
\begin{equation}
K\frac{\partial T}{\partial r} = F_{\rm Sun} + F_{\rm Sat} -\epsilon_{\rm IR} \sigma T^4 \left(1-\frac{\Omega_{\rm p}}{4\pi}\right), \label{eq:bc}
\end{equation}
where $F_{\rm Sun}$ and  $F_{\rm Sat}$ are the fluxes from the Sun and Saturn, respectively, $\sigma_{\rm SB}$ is the Stefan-Boltzmann constant, 
and $\Omega_{\rm p}$ is the solid angle of surrounding particles.
The third term in the right hand side of Eq.~(\ref{eq:bc}) is the radiation flux from the particle, and a part of the emission
is blocked by surrounding particles that are assumed to have the same temperature as the subject particle.
We adopt $\Omega_{\rm p} =  6[1-\exp(-\tau)]$ (Ferrari et al., 2005), where $\tau$ is the ring optical depth.

The solar flux is given by 
\begin{equation}
F_{\rm Sun} = \frac{1-A}{f} S(B',\tau) \left(\frac{r_{\rm S}}{{\rm 1 AU}}\right)^{-2} F_{\odot}, \label{eq:fsun}
\end{equation}
where $A$ is the albedo of the particle (either $A_{\rm lit}$ or $A_{\rm unlit}$ defined below), $f$ is the spin factor, 
$F_{\odot}$ = 1370 Wm$^{-2}$ is the solar flux at 1AU.
We introduce two albedos: the lit-face albedo $A_{\rm lit}$ that is used when the face is directly illuminated by the Sun
and the another one for the unlit face $A_{\rm unlit}$. It is expected that $A_{\rm lit}$ is close to the bolometric Bond albedo of the particle.
On the other hand,  $A_{\rm unlit}$ includes the effect due to vertical heat transport that decreases with increasing $\tau$. Thus,
it is expected that $A_{\rm unlit}$ is larger than $A_{\rm lit}$ as also shown by Flandes et al. (2010). 
The model assumes that the timescale of vertical heat transport is short compared with the orbital period of Saturn, $T_{\rm S}$. 
This is a reasonable assumption if vertical heat transport is taking place by particle motion (Morishima et al., 2010; Pilorz et al., 2014). 
The spin factor $f$ is 2 for slow rotators and 4 for fast rotators.
We adopt $f =3$ that is suggested from equinox modeling (Spilker et al., 2013).
In Eq.~(\ref{eq:fsun}), $S$ represents the factor due to shadowing of surrounding particles and we use the approximate formula 
given by Altobelli et al. (2008): 
\begin{equation}
S(B',\tau) = \frac{\sin{|B'|}}{1-\exp{(-\tau)}}\left[1-\exp\left(-\frac{\tau}{\sin{|B'|}}\right)\right].
\end{equation}

The Saturn flux is given by 
\begin{equation}
F_{\rm Sat} = \epsilon_{\rm IR} \sigma T_{\rm eq}^4 \left(1-\frac{\Omega_{\rm p}}{4\pi}\right),
\end{equation}
where $T_{\rm eq}$ is the equilibrium equinox temperature due to the Saturn flux only 
(note that $T = T_{\rm eq}$ if $F_{\rm Sun}$ = $K$ = 0). 
As stated above, we take $T_{\rm eq}$ from our radiative transfer calculations shown in Section~3. 
We fix $\epsilon_{\rm IR} = 1$ in the seasonal model. 
Note that $\epsilon_{\rm IR}$ is not  a free parameter as scattered thermal light is not taken into account in the model; 
dividing Eq.~(\ref{eq:bc}) by $\epsilon_{\rm IR}$, the meaningful parameters are found to be $K/\epsilon_{\rm IR}$ and $(1-A)/(f\epsilon_{\rm IR})$, 
but not $\epsilon_{\rm IR}$ itself.

The time in the thermal diffusion equation can be normalized by the inverse of the mean motion of Saturn, $\omega_{\rm S}^{-1}$. 
The length can be normalized by 
the seasonal thermal skin depth $\ell$ given by 
\begin{eqnarray}
\ell &=& \sqrt{\frac{K}{\rho C \omega_{\rm S}}} 
= \frac{\Gamma}{\rho C \sqrt{\omega_{\rm S}}}
\nonumber \\ 
&\simeq& 
0.60
\left(\frac{\Gamma}{10 \hspace{0.3em}{\rm J \hspace{0.1em} m^{-2} K^{-1} s^{-1/2}}}\right)
\left(\frac{\rho}{450\hspace{0.3em}{\rm kg \hspace{0.1em}m^{-3}}}\right)^{-1}
{\rm m},
\label{eq:skin} 
\end{eqnarray} 
where $\Gamma = \sqrt{\rho C K}$ is the thermal inertia.
This length is a few order of magnitude larger than the diurnal skin depth, $\sim$ 1 mm (Morishima et al., 2009),
and comparable to the ring particle size suggested from occultations (French and Nicholson, 2000; Cuzzi et al., 2009).
After the normalization of Eq.~(\ref{eq:difeq}), $\Gamma$ appears as a single parameter rather than 
$\rho$, $C$, and $K$, independently. 
If $\ell \gg R$, the thermal relaxation time (or the cooling time) of the particle is given by 
\begin{eqnarray}
\tau_{\rm rel1} &=& \frac{(4\pi/3) R^3 \rho C T}{4\pi R^2 \epsilon_{\rm IR} \sigma T^4} = \frac{R \rho C}{3\epsilon \sigma T^3}\nonumber \\
&\simeq& 110
\left(\frac{R}{1\hspace{0.3em}{\rm m}}\right)
\left(\frac{\rho}{450\hspace{0.3em}{\rm kg \hspace{0.1em} m^{-3}}}\right)
\left(\frac{T}{50 \hspace{0.3em} {\rm K}}\right)^{-3} {\rm days}. \label{eq:rel1}
\end{eqnarray} 
Thus, the equinox temperature sensitively depends on $R\rho C$.
On the other hand, if $\ell \ll R$, the thermal relaxation time is given by 
\begin{eqnarray}
\tau_{\rm rel2} &=& \frac{4\pi R^2 \ell \rho C T }{4\pi R^2 \epsilon_{\rm IR} \sigma T^4} = \frac{\Gamma}{\epsilon \sigma T^3 \sqrt{\omega_{\rm S}}}\nonumber \\
&\simeq& 198
\left(\frac{\Gamma}{10 \hspace{0.3em}{\rm J \hspace{0.1em} m^{-2} K^{-1} s^{-1/2}}}\right)
\left(\frac{T}{50 \hspace{0.3em} {\rm K}}\right)^{-3} {\rm days}. \label{eq:rel2}
\end{eqnarray} 
Thus, the equinox temperature sensitively depends on $\Gamma$.
If $R \sim \ell$, the thermal relaxation time takes an intermediate value between $\tau_{\rm rel1}$ and $\tau_{\rm rel2}$ and 
the equinox temperature depends on both $R\rho C$ and $\Gamma$.

\subsubsection{Particle internal structure models}

\begin{figure}
\begin{center}
\includegraphics[width=1.\textwidth]{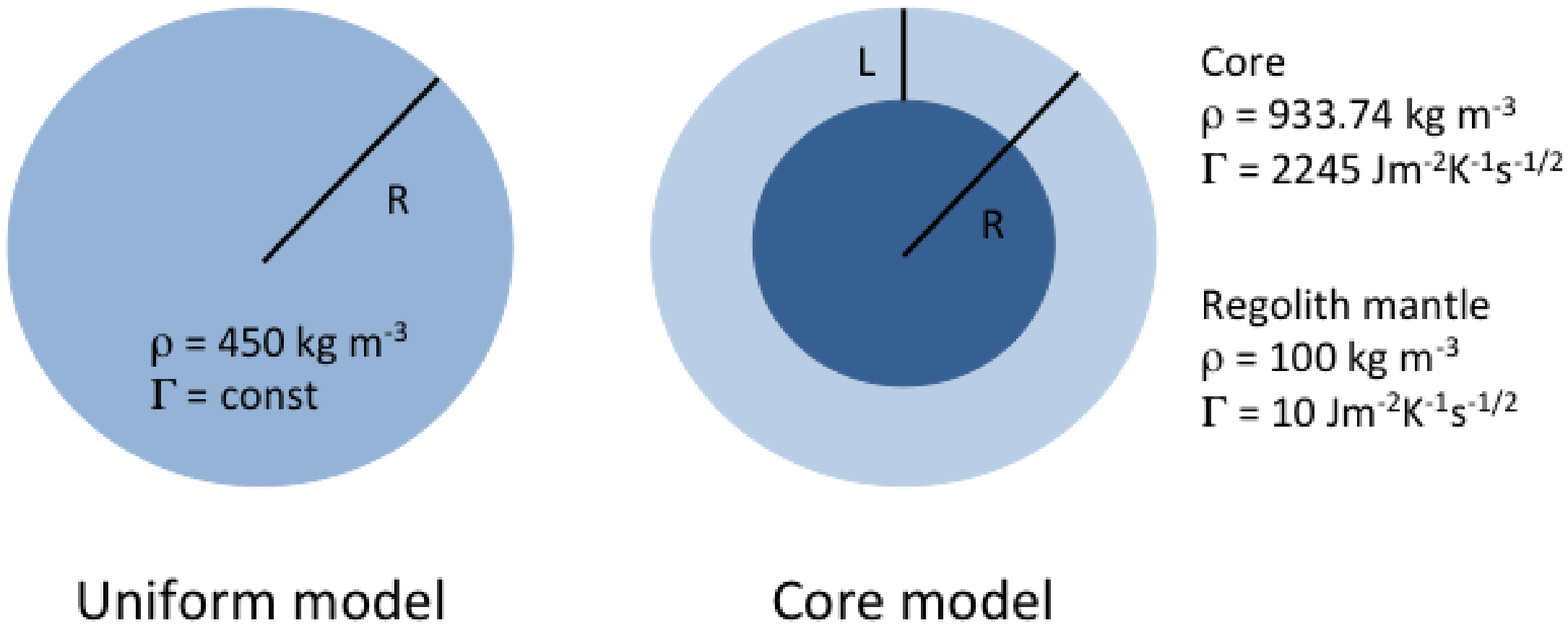}
\end{center}

Fig.~9. Schematic illustration of two internal structure models.

\end{figure}

For the interior of the particle, we consider two models (Fig.~9).
The first model,  called the uniform model, adopts uniform $\Gamma$ and $\rho$ with depth. 
We adopt $\rho = 450\hspace{0.3em}{\rm kg \hspace{0.1em} m^{-3}}$. This value is suggested from 
photometric azimuthal brightness asymmetry (Salo et al., 2004, French et al., 2007).
The parameters to be fitted are $\Gamma$, $R$, $A_{\rm lit}$, and $A_{\rm unlit}$.  
For simplicity, we consider only single-sized particles whereas actual ring particles have a size distribution (Cuzzi et al., 2009).
The inner boundary for Eq.~(\ref{eq:difeq}) is set to be 
a larger one of $0$ and $R-7\ell$, and the boundary condition there is given as 
\begin{equation}
\left.\frac{\partial T}{\partial r}\right|_{\rm inner } = 0.
\end{equation}

In the second model,  called the core model, we consider a dense core with a radius $R_{\rm c}$ covered by a fluffy regolith mantle
with a thickness of $L = (R-R_{\rm c})$.
We assume that the core is made of ice Ih with no porosity and 
adopt the properties at 50 K: 
the internal density $\rho_{\rm c} = 933.74 \hspace{0.3em}{\rm kg \hspace{0.1em} m^{-3}}$ (Feistel and Wagner, 2006) 
and the thermal conductivity $K_{\rm c}$ = 12 W m$^{-1}$ K$^{-1}$ (Andersson and Inaba, 2005).
These values give the thermal inertia of the core as
$\Gamma_{\rm c} = 2245 \hspace{0.3em} {\rm J \hspace{0.1em} m^{-2} K^{-1} s^{-1/2}}$.
The regolith mantle is assumed to have $\rho = 100\hspace{0.3em}{\rm kg \hspace{0.1em} m^{-3}}$
and $\Gamma   = 10$ ${\rm J \hspace{0.1em} m^{-2} K^{-1} s^{-1/2}}$; 
the value of $\Gamma$ is those estimated from eclipse cooling (Morishima et al. 2011, 2014) and 
$\rho$ is those indicated from the relationship between the porosity and the thermal conductivity (see Fig.~14 for details). 
In the core model, $R$ and $L/R$
are varied as fitting parameters. We explore only in the range of 0.05 $\le L/R \le$ 0.5, as 
a large(small) $L/R$ gives a mean internal density that is too low(high). 
We use $A_{\rm lit}$ and $A_{\rm unlit}$ from the uniform model. 
Since the core's $K$ is large enough, the thermal diffusion time scale for a meter-size core 
is short enough (less than a day).
Thus, the spatial temperature variation inside the 
core can be ignored and the core can be well represented by a single grid in numerical calculations. 
The equation of the core temperature change is given by 
\begin{equation}
\frac{R_{\rm c} \rho C}{3} \frac{\partial T}{\partial t} = 
K _{\rm b} \left.\frac{\partial T}{\partial r}\right|_{r=R_{\rm c}} , \label{eq:core}
\end{equation}
where the thermal conductivity, $K_{\rm b}$, at the core-mantle boundary is set to be $2(K_{\rm c}^{-1} + K^{-1})^{-1} \simeq 2K$ (Shoshany et al., 2002).

For the uniform model, the particle or the surface portion is divided into 100 layers with an identical thickness, and 
the temperature change is numerically calculated using  the Crank-Nicholson method (Press \textit{et al.}, 1986; Chap.~19.2).
After convergence tests, we adopt the timestep of $10^{-6} T_{\rm S}$.
This very small timestep is required because the solar flux changes very rapidly on a short timescale  around the equinox.
For the core model, the regolith mantle is divided into multiple layers. 
It is found that numerical instabilities occur if the thickness of each layer is too small as compared with the skin depth. 
After some experiments, we adopt the number of layers, $N_{\rm L}$, as follows:
\begin{equation}
N_{\rm L} = {\rm min}(2200\frac{L}{\ell},100).
\end{equation}
%For the smallest $L$  (0.625 cm) in our simulations,
%the regolith mantle is divided into only five layers. Nevertheless, numerical errors are found to be small enough, by
%comparing with simulations with finer grid sizes and smaller timestep sizes.  

In the seasonal model, 
we again use the optical depths from Voyager PPS (Esposito et al., 1983) 
smoothed over each radial bin, avoiding the Encke division. 
We also used some of profiles taken by Cassini UVIS (Colwell et al., 2010), but
we did not see any significant differences in the fitted parameters. 
Thus, those results will not be presented.

\subsection{Fitting procedures} 
For model fitting, we divide the observed data into four portions: 
(1) the southern face data before the equinox, (2) the southern face data after the equinox,
(3)  the northern face data before the equinox, and (4) the northern face data after the equinox.
The total number of scans for the portion (1) is defined to be $n_{\rm s1,max}$, and the numbers for 
other portions are defined as well. 
The $n_{\rm s1}$-th scan in the portion (1) contains $n_{\rm f1,max}(n_{\rm s1})$ temperatures in a designated radial bin 
of the width of 1,000 km.
The mean temperature and the standard deviation of the $n_{\rm s1}$-th scan are 
defined to be $T_{\rm obs1} (n_{\rm s1})$ and $\sigma_{\rm obs1}(n_{\rm s1})$.
The reduced chi-square for the portion (1), $\chi_1^2$,  is then calculated as 
\begin{eqnarray}
&& \chi_1^2 (\Gamma, R, A_{\rm lit},A_{\rm unlit}) = \frac{1}{n_{\rm s1,max}-M}   \times \nonumber \\ 
&& \sum_{n_{\rm s1}=1}^{n_{\rm s1,max}}w(n_{\rm s1})\left(\frac{T_{\rm obs1}(n_{\rm s1}) - T_{\rm sim1}(n_{\rm s1},\Gamma, R, A_{\rm lit},A_{\rm unlit})}{\sigma_{\rm obs1}(n_{\rm s1})}\right)^2, \label{eq:ind}
\end{eqnarray}
where $M = 4$ is the number of the parameters to be fitted, $T_{\rm sim1}$ is the modeled temperature, and $w(n_{\rm s1})$ is the weight given to each scan. 
After some experiments, we determined to adopt the following form for the weight:
\begin{equation}
w(n_{\rm s1}) = c_1n_{\rm f1,max}(n_{\rm s1}) \exp{\left(- \frac{|B'|}{\rm 3 \hspace{0.2em} deg}\right)} \label{eq:wei},
\end{equation}
where $B'$ is the solar elevation angle at the time when the scan was taken and the coefficient $c_1$
is derived by normalization of the weight as
\begin{equation}
 \sum_{n_{\rm s1}=1}^{n_{\rm s1,max}}w(n_{\rm s1}) = 1.
\end{equation}
Eq.~(\ref{eq:wei}) means that  more weight is given to data near the equinox. 
The temperature  at high $|B'|$ is almost the equilibrium temperature obtained by ignoring the time dependence,
and is primarily adjusted by the albedos $A_{\rm lit}$ and $A_{\rm unlit}$.
Therefore, we cannot well constrain $\Gamma$ and $R$ if we weight all the data equally.

The reduced chi-squares for the portions (2)-(4) are calculated in a similar manner.
Then, the reduced chi-square for the uniform model is simply defined by averaging the four values as
\begin{equation}
 \chi_{\rm A}^2 = \frac{1}{4} \left(\chi_1^2 +  \chi_2^2 +  \chi_3^2 +  \chi_4^2 \right). 
\end{equation}
A set of the best-fit parameters are obtained at the chi-square minimum,  $\chi_{\rm A,min}^2$, in the four-dimensional 
parameter space.
The error bars are estimated from the maximum and minimum values of the parameters on the contour of $\chi_{\rm A}^2/\chi_{\rm A,min}^2 = 2$.   
A similar procedure gives the reduced chi-square, $\chi_{\rm B}^2$, for the core model and  
the best-fit parameters are obtained at the chi-square minimum, $\chi_{\rm B,min}^2$.

% (ranges of parameters)
 
\subsection{Uniform model results}
\begin{figure}

\begin{center}
\includegraphics[width=.6\textwidth]{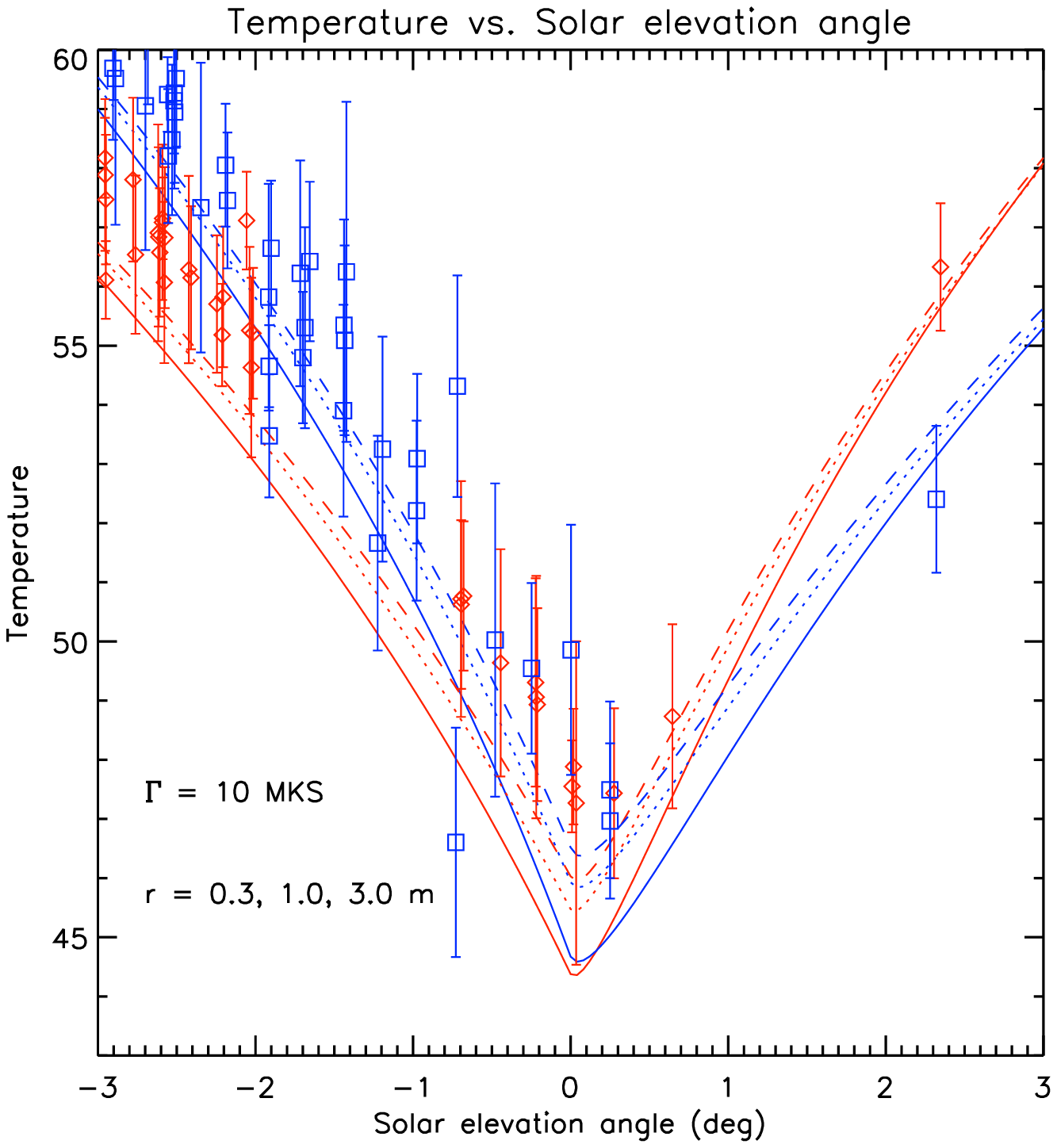}
\includegraphics[width=.6\textwidth]{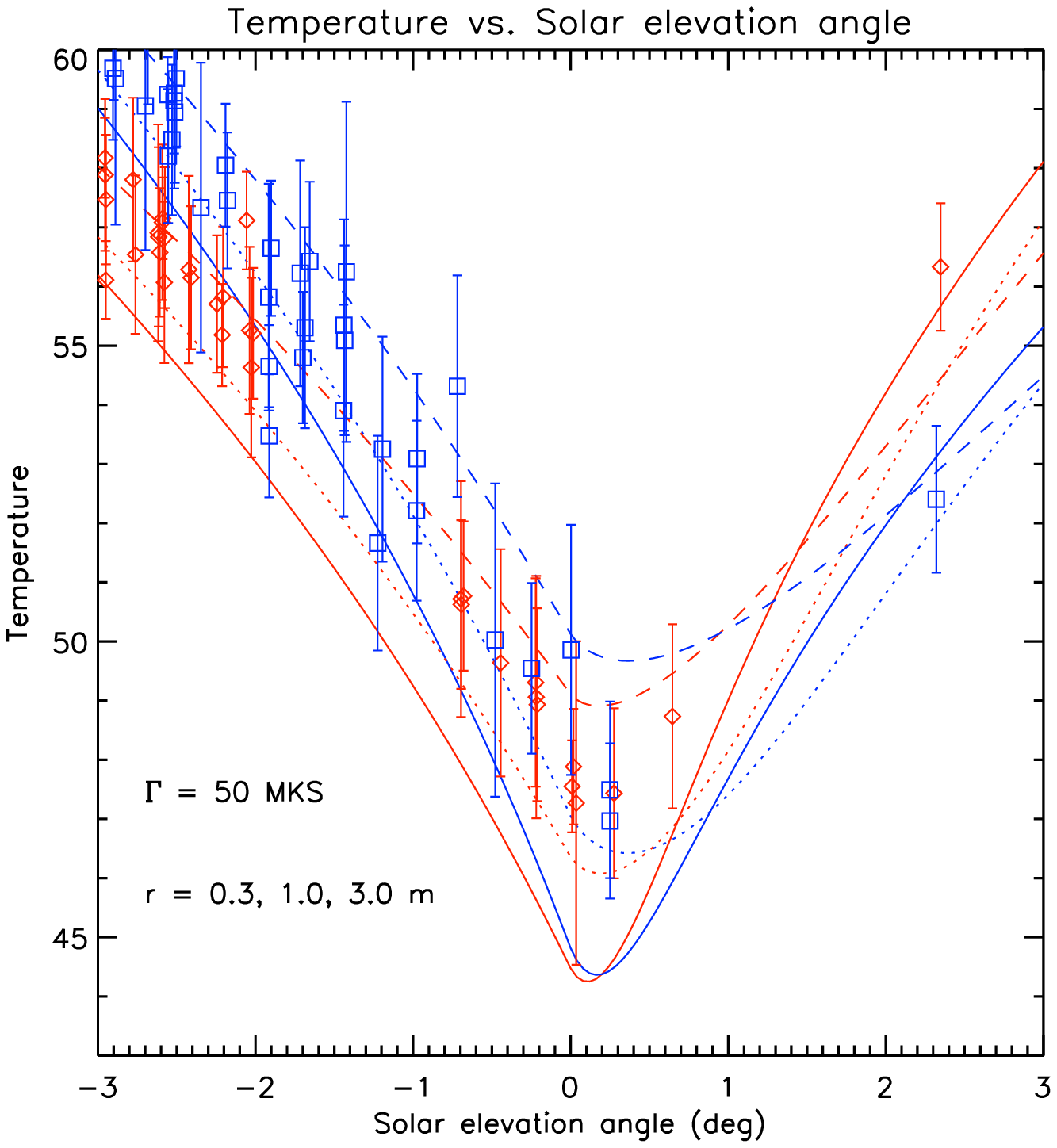}
\end{center}

Fig.~10. 
Examples of seasonal temperature variations for the uniform model at $129,000$ km. 
The observational data are the same as those in Fig.~7.
The upper and lower panels show the cases of $\Gamma = 10$ and 50 Jm$^{-2}$K$^{-1}$s$^{-1/2}$.
The results of three different radii $R = 0.3$(solid), 1.0(dotted), and 3.0 m (dashed) are shown 
for the southern (blue) and northern (red) faces.

\end{figure}

\begin{figure}
\begin{center}

\includegraphics[width=.6\textwidth]{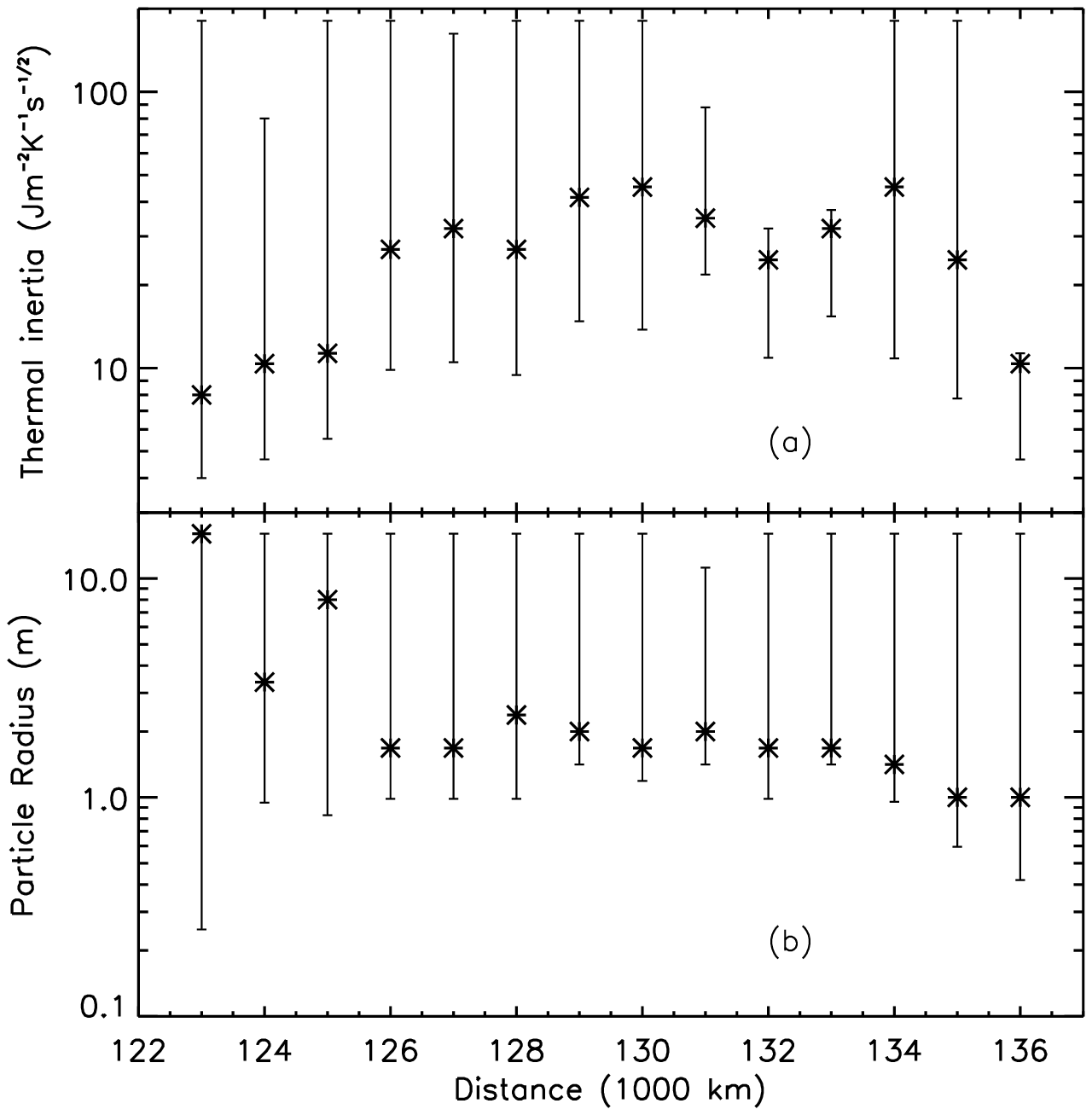}
\includegraphics[width=.6\textwidth]{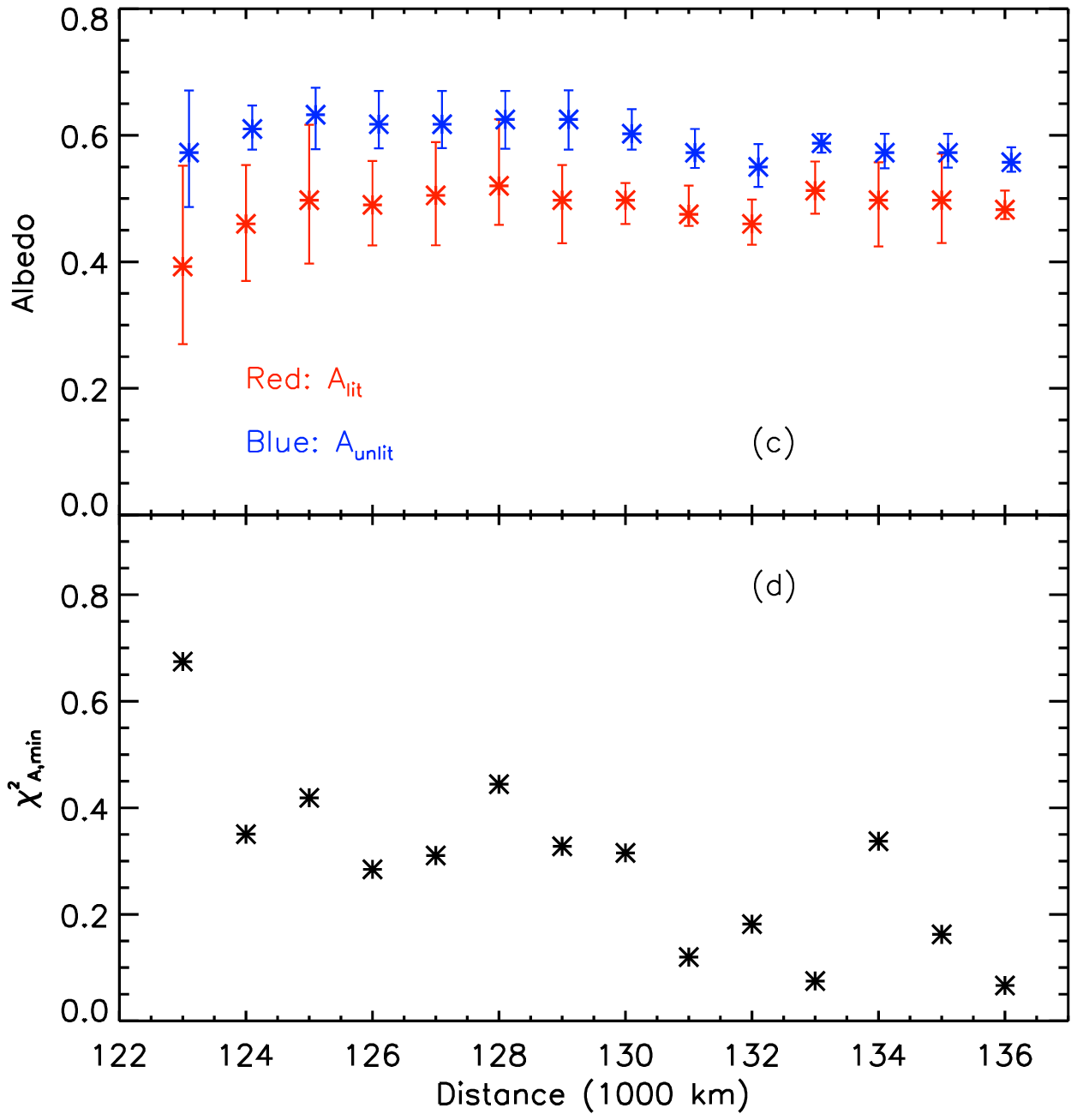}
\end{center}

Fig.~11. 
Estimated parameters for the uniform model: 
(a) thermal inertia $\Gamma$, (b) particle radius $R$, (c) particle albedos $A_{\rm lit}$ and $A_{\rm unlit}$,
and (d) goodness of fits $\chi_{\rm A,min}^2$.
The ranges of $\Gamma$ and $R$ are limited a priori in data fits as 4 Jm$^{-2}$K$^{-1}$s$^{-1/2}$ 
$\le \Gamma \le $ 181 Jm$^{-2}$K$^{-1}$s$^{-1/2}$ and 
0.125 m $\le R \le$ 16 m.

\end{figure}

\begin{figure}
\begin{center}

\includegraphics[width=.49\textwidth]{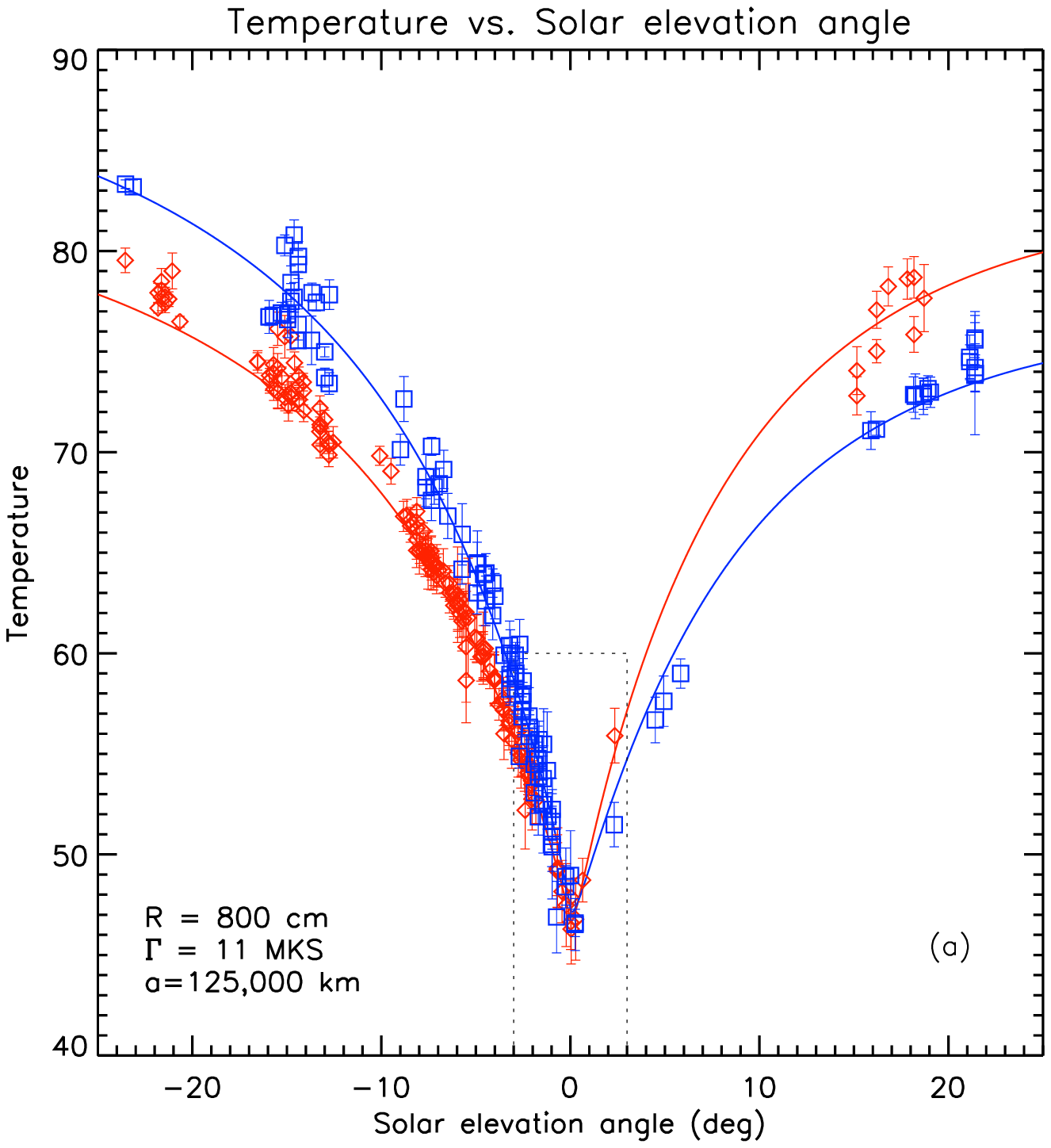}
\includegraphics[width=.49\textwidth]{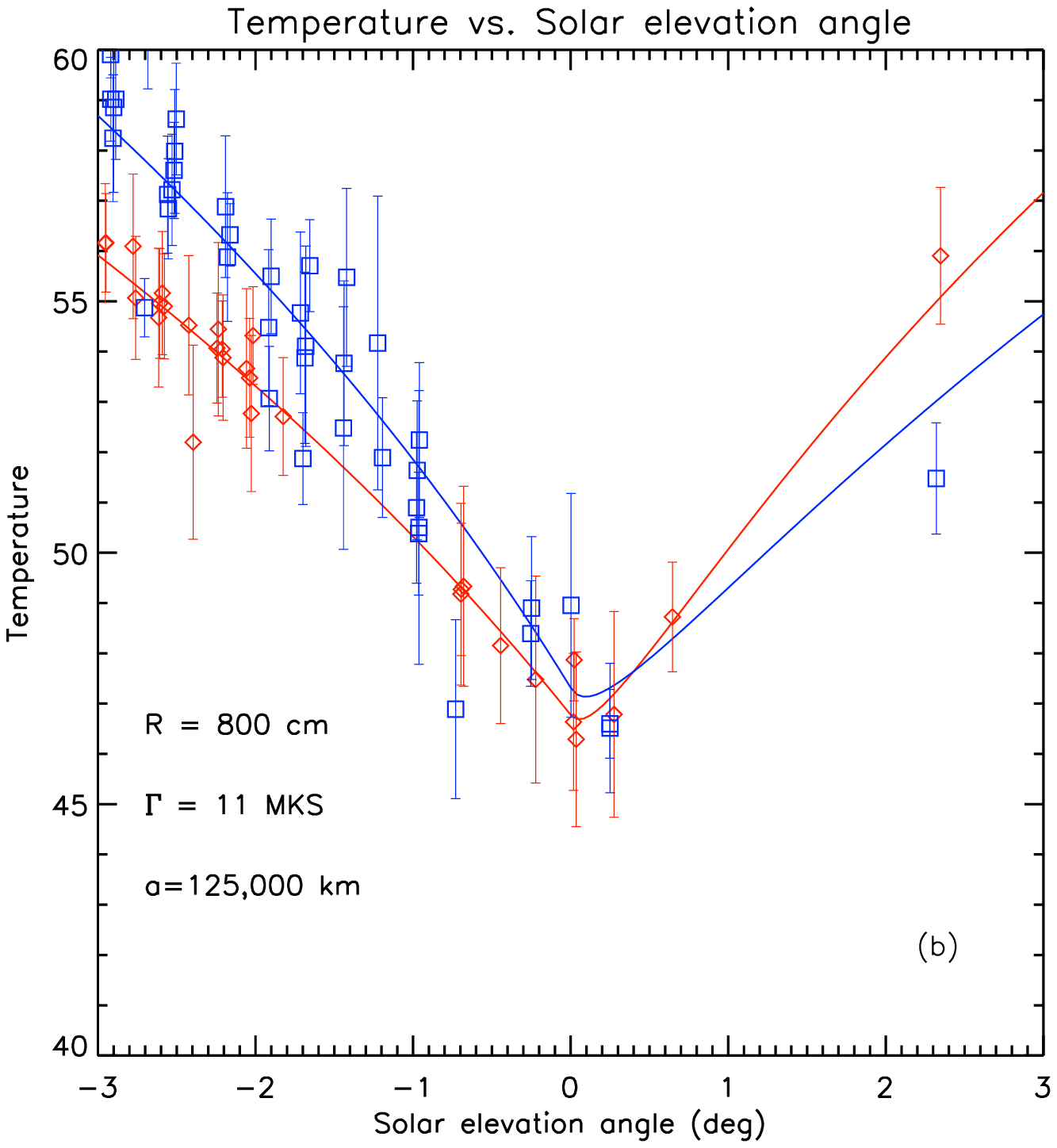}
\includegraphics[width=.65\textwidth]{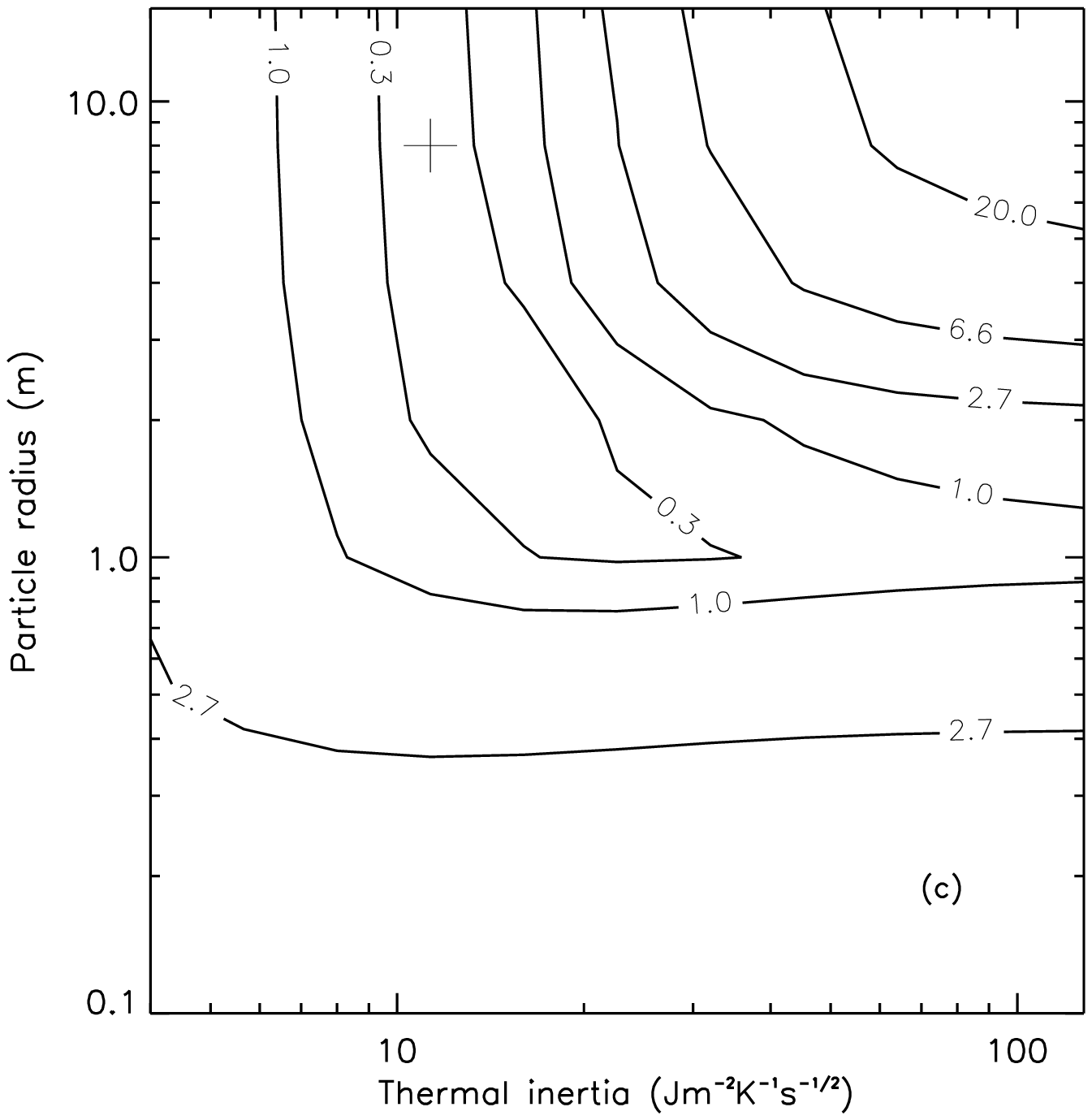}
\end{center}

Fig.~12. 
Data fitting with the uniform model at $125,000$ km.
(a) Best-fit temperature curves as a function of $B'$.
The northern and southern face temperatures are represented by red and blue colors, respectively.
(b) Enlargement of the region surrounded by dotted lines in (a).
(c) Contour of $\chi_{\rm A}^2/\chi_{\rm A,min}^2-1$ on the $\Gamma$ vs. $R$ plane sliced at the best-fit
$A_{\rm lit}$ and $A_{\rm unlit}$. The point of the best-fit $\Gamma$ and $R$ is shown by the plus mark.

\end{figure}

\begin{figure}
\begin{center}

\includegraphics[width=.49\textwidth]{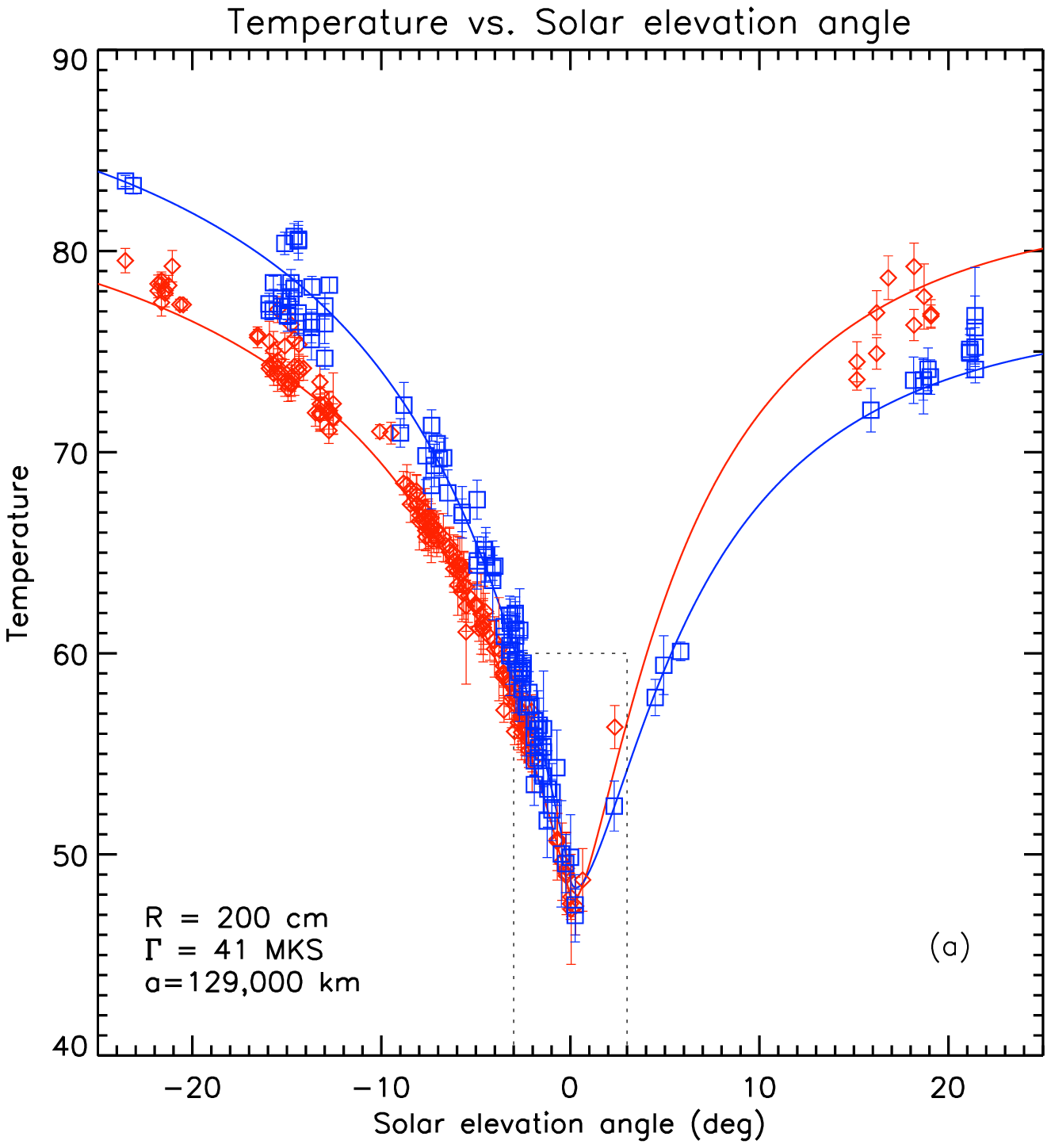}
\includegraphics[width=.49\textwidth]{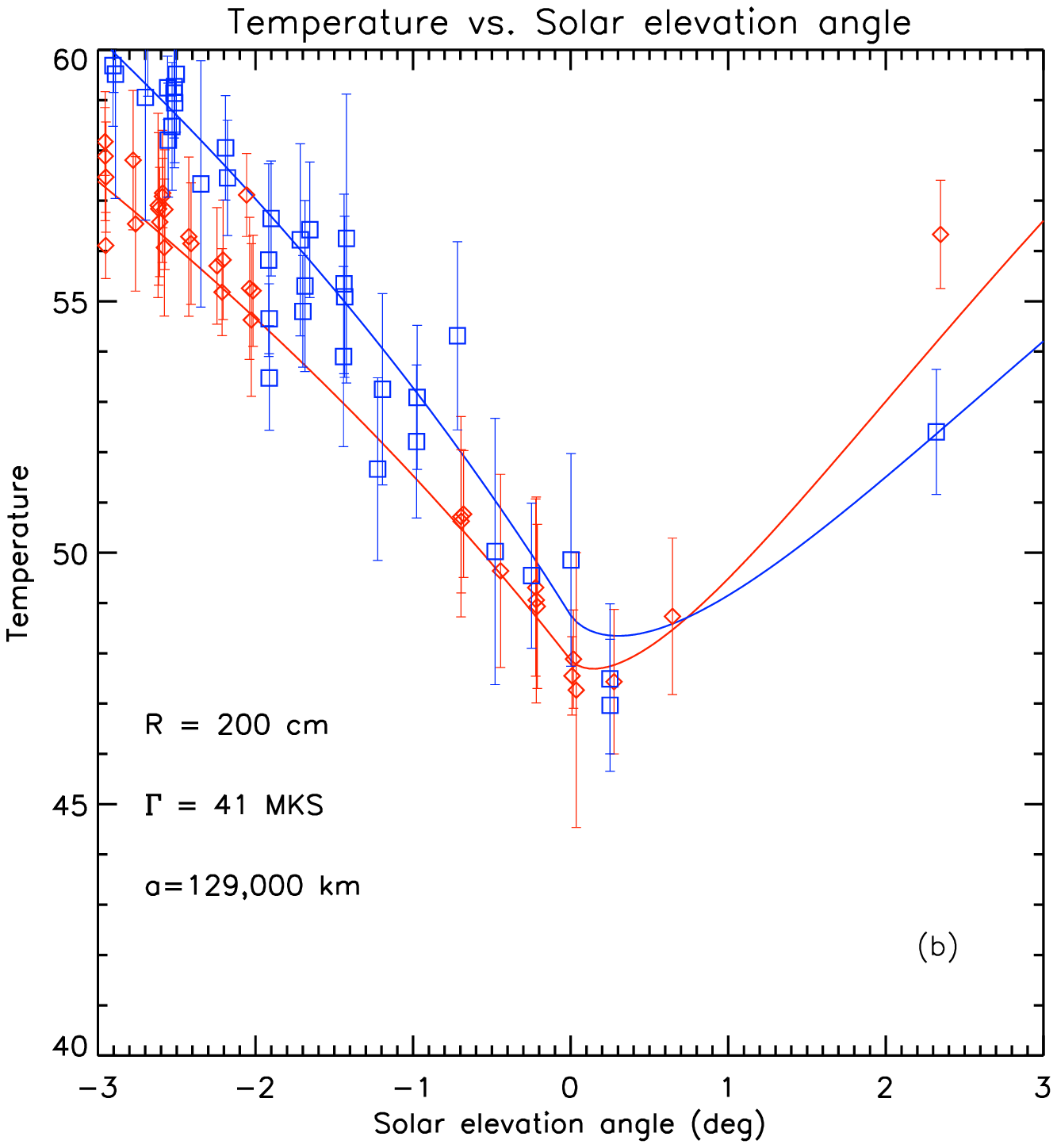}
\includegraphics[width=.65\textwidth]{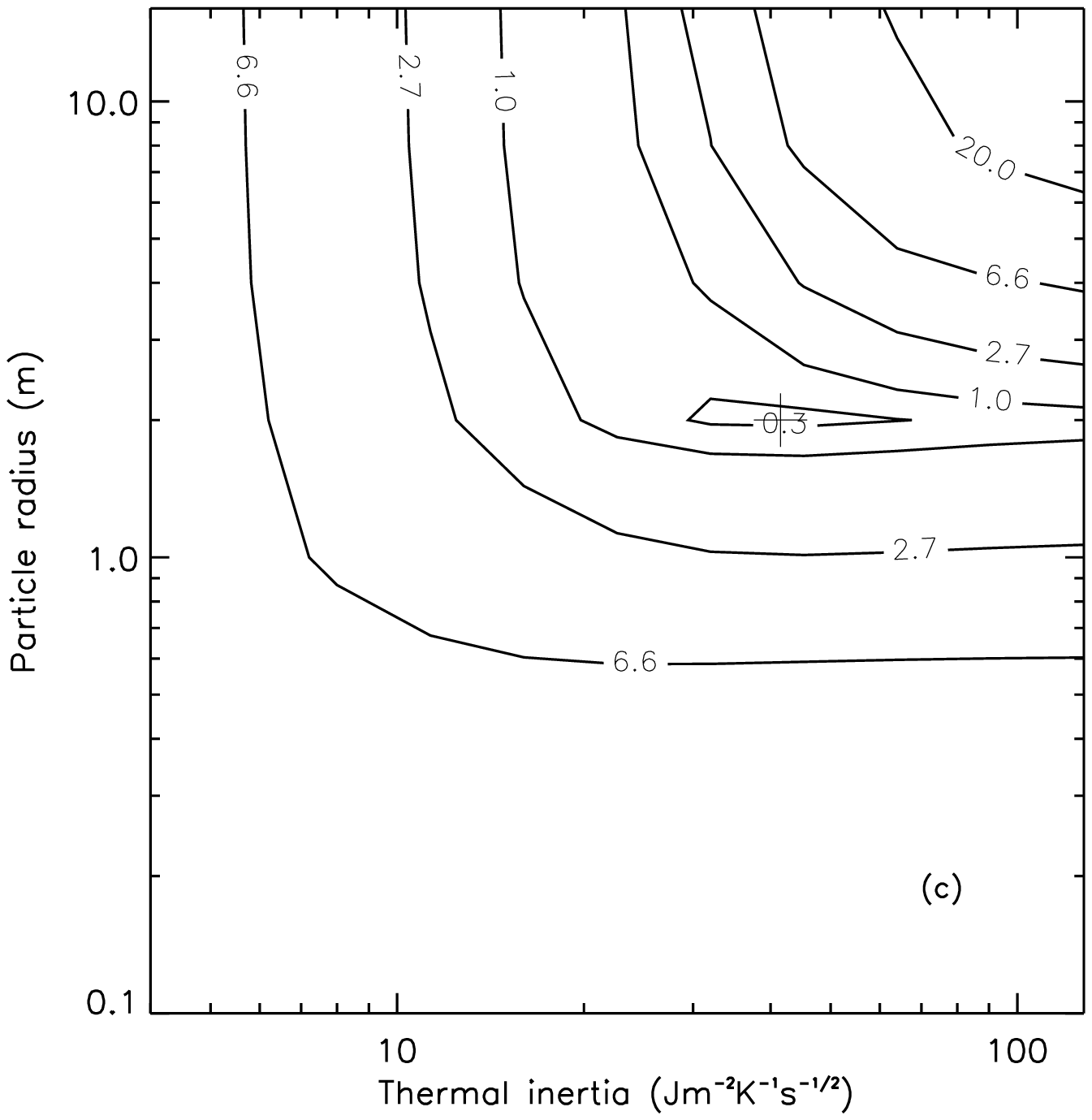}

\end{center}

Fig.13. Same as Fig.~12, but at $a =129,000$ km.

\end{figure}

Before discussing the best-fit parameters,  we show how  seasonal temperature curves change with 
different input parameters. Figure~10 shows the observed and modeled temperatures near the equinox at 129,000 km. 
In the model, the particle size and the thermal inertia are varied while the best-fit values of the albedos, $A_{\rm lit} = 0.498$ and $A_{\rm lit} = 0.625$, from Fig.~11
are used.  The upper panel shows the case of $\Gamma =10$ ${\rm J \hspace{0.1em} m^{-2} K^{-1} s^{-1/2}}$. The seasonal thermal skin depth $\ell$  
is 0.60 m in this case.  The seasonal temperature variation is almost independent of $R$ as long as $R  > \ell$. 
For the case of $R = 0.3$ m, the equinox temperature is slightly lower than large $R$ cases. 
Both the lit and unlit face temperatures are nearly symmetric with respect to the equinox, 
as the thermal relaxation time, $\tau_{\rm rel2}$[Eq.~(\ref{eq:rel2})] (= 198 days), is short enough.
The timescale $\tau_{\rm rel2}$ roughly coincides with the duration of $|B'| < 1^{\circ}$, and 
 it is expected that the ring temperature is nearly the equilibrium temperature for $|B'| > 1^{\circ}$ in this case.
The lower panel shows the case of $\Gamma =50$ ${\rm J \hspace{0.1em} m^{-2} K^{-1} s^{-1/2}}$ and this case gives $\ell = 3.0$ m.
In this case, the equinox temperature depends on $R$, as $R \le \ell $.
The lit and unlit face temperatures are asymmetric with respect to the equinox particularly for $R = 1.0$ and 3.0 m; the temperature after the equinox is 
lower than that before the equinox at the same $|B'|$.
The thermal relaxation time, $\tau_{\rm rel2}$ is 993 days, and this is roughly the duration of $|B'| < 6-7^{\circ}$. Indeed,
we see convergence of the temperature curves for three different values of $R$ when $|B'| > 6-7^{\circ}$ (outside of the figure).
For $R = 0.3$ m,  the temperature curves show only a small dependence on $\Gamma$ and
this indicates that the thermal relaxation time is given by $\tau_{\rm rel1}$ (33 days).
The modeled equinox temperatures are lower than the observed ones if $\Gamma = 10$ ${\rm J \hspace{0.1em} m^{-2} K^{-1} s^{-1/2}}$ or 
$R = 0.3$ m. Thus, $\Gamma$ and $R$ are expected to be larger than those values at 129,000 km.

Figure~11 shows the best-fit values of $\Gamma$, $R$, $A_{\rm lit}$, and $A_{\rm unlit}$. The reduced chi-square minimum, $\chi_{\rm min, A}^2$ is also shown.
The thermal inertia $\Gamma$ is typically 30-50 ${\rm J \hspace{0.1em} m^{-2} K^{-1} s^{-1/2}}$ in the middle A ring, while 
it is $\sim$ 10 ${\rm J \hspace{0.1em} m^{-2} K^{-1} s^{-1/2}}$ in the inner and outermost A ring. 
The latter value is close to the diurnal thermal inertia derived from eclipse temperatures (Morishima et al., 2011, 2014). 
The particle size is 1 -2 m in most of saturnocentric distances but larger values are seen in the inner A ring. 
It is found that the error bars of $\Gamma$ and $R$ are very large and the upper limits of these parameters are not usually well constrained except 
for some cases such as $\Gamma$ at 136,000 km. 
The value of $A_{\rm lit}$ is about 0.4-0.5 (note again that we adopt the spin factor $f = 3$), consistent with those estimated by Morishima et al. (2010). 
The value of $A_{\rm unlit}$ is about 0.6. The ratio of $A_{\rm unlit}$ to $A_{\rm lit}$ is consistent with those of Flandes et al. (2010).
In all radial locations, $\chi_{\rm A,min}^2 < 1$, indicating that the fits are good. 

Examples of the best-fit temperature curves are shown in Figs.~12 ($a=$ 125,000 km) and 13 ($a=$ 129,000 km).
The fits are found to be good for the entire range of $B'$. The temperature asymmetry near the equinox is caused by the effect of incomplete cooling down as 
we discussed, while the asymmetry at high $|B'|$ is caused by different saturnocentric distances.
Figures~12c and 13c shows the contours of  $\chi_{\rm A}^2/\chi_{\rm A,min}^2 - 1$ on the plane of $\Gamma$ vs. $R$ sliced at the best-fit values of $A_{\rm lit}$ and $A_{\rm unlit}$.
As shown in Fig.~11, $\Gamma$ and $R$ are not well constrained independently. 
However, the region of small $\chi_{\rm A}^2$ values is confined in a narrow band on the plane of $\Gamma$ vs. $R$. 
We can clearly exclude particle with both high $\Gamma$ and $R$. 
The lower limits of $\Gamma$ and  $R$ are also well constrained. 
The contour for $a$ = 129,000 km has a band of low $\chi_{\rm A}^2$ which is shifted toward 
a upper-right position compared with that for $a$ = 125,000 km.  
This means that at least one of $\Gamma$ or $R$ needs to be larger at $a$ = 129,000 km than $a = 125,000$ km.
French and Nicholson (2000) showed that the size distribution of ring particles is similar across the A ring inside 
the Encke gap ($a$= 133,589 km). Thus,  the radial variation of $R$ is probably small ($R$ is likely to be 1-2 m even for the inner A ring),
whereas  $\Gamma$ should be relatively large in the middle A ring 
compared with the inner A ring (see Section~5 for more discussions). 

\subsubsection{Expected regolith density}
\begin{figure}

\begin{center}
\includegraphics[width=.75\textwidth]{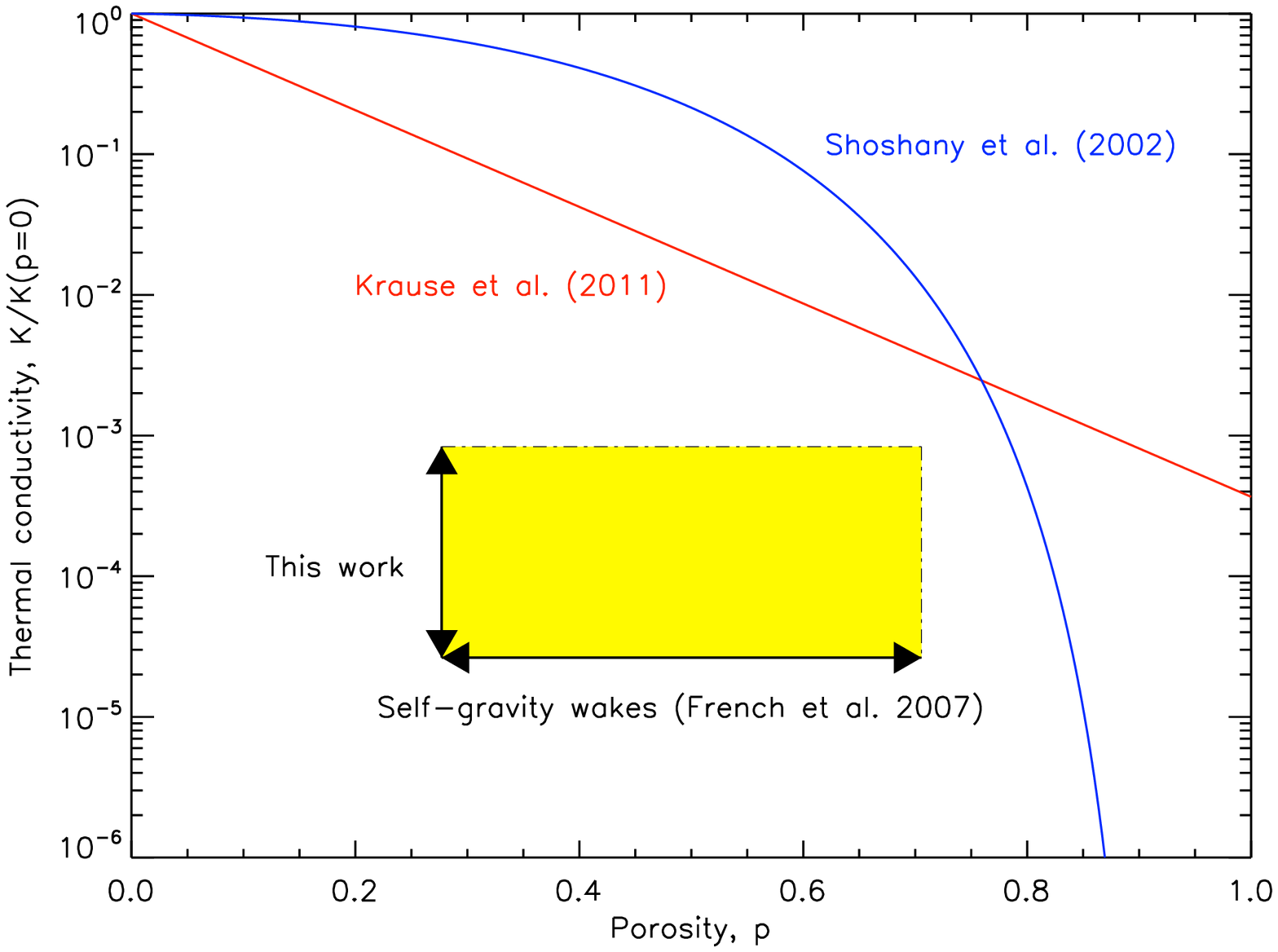}
\end{center}

Fig.~14.
The thermal conductivity $K$ as a function of porosity, $p$.  The conductivity is normalized by its value at $p=0$.
The numerical model of Shoshany et al. (2002)  gives a fitting curve, $K/K(p=0) = (1-p_0/0.7)^{(4.1p_0+0.22)n}$, where $p_0 = 1-(1-p)^{1/n}$. 
In their model, the ratio of the largest grain size to the smallest grain size is given by $50^{(n-1)}$ and we adopt $n=2$. 
Krause et al. (2011) conducted heat conduction experiments 
using silicon beads and found the relationship $K/K(p=0) = \exp{(-7.91p)}$ (their maximum $p$ is 0.85). 
The yellow rectangle indicates the estimated ranges of $K$ and $p$ for the A ring particles. 
For the range of $p$, we take $\rho = 270$-675 kg m$^{-3}$ (the upper and lower limits are not strict) 
from Fig.~23 of French et al. (2007) and assume that  ring particles are pure water ice; $\rho = 933.74$ kg m$^{-3}$ at $p$ = 0. 
The range of $K$ is from this study.

\end{figure}

The seasonal thermal inertia estimated in our model (Fig.~11) is found to be comparable to or somewhat larger 
than the diurnal thermal inertia. The seasonal thermal inertia is still much lower than that for rigid water ice, 
$\sim 2200$ ${\rm J \hspace{0.1em} m^{-2} K^{-1} s^{-1/2}}$,  indicating that regolith of ring particles is likely to be very fluffy down to their deep interiors. 
Many studies have been made of the relationship between the thermal conductivity, $K$,  and the porosity, $p$. 
The exact relationship varies from author to author, but in principle $K$ significantly decreases with increasing $p$.   
Figure~14 shows two examples that are likely to be relevant to Saturn's rings; 
one derived from numerical simulations of Shoshany et al. (2002) and another 
one derived from laboratory experiments of Krause et al. (2011).
The range of $K(p)/K(0)$ from our model results is about $10^{-5}$ - $10^{-3}$ and is shown by the arrow in Fig.~14, where we use 
the values of $\rho$ and $C$ assumed in our model in the conversion from $\Gamma$ to $K$.
Although the two curves of $K$ differ from each other, both indicate that a high porosity ($p > 0.8$) is necessary 
to give $K$ as low as those estimated from our model.  
If ring particles mostly consist of water ice (Cuzzi et al., 2009), 
this means that $\rho \sim 200$ kg m$^{-3}$ or even less.
Probably, ring particle regolith is as fluffy as fresh snow on the Earth ($\rho \sim 100$ kg m$^{-3}$; Judson and Doesken, 2000), 
as the regolith is likely to be impact ejecta that gently accumulated on particles 
(Cuzzi and Estrada, 1998; Elliott and Esposito, 2011, Tiscareno et al., 2013; Schmidt and Tiscareno, 2013). 

However, it is known that $\rho$ of A ring particles should not be that low, because self-gravity 
wakes are observed. French et al. (2007) compared the azimuthal brightness asymmetry observed by the Hubble Space Telescope
with those from their $N$-body ray-tracing simulations and found that the observed amplitude of azimuthal asymmetry is reasonably reproduced 
if $\rho \sim 450$ kg m$^{-3}$ (the value we assumed in the uniform model). Thus, the uniform model is not self-consistent
as the estimated $K$ is too low for the suggested value of $\rho$ or $p$ (Fig.~14).
To resolve this issue, in the next section, we introduce the core model that allows moderately high mean internal density and 
very fluffy surface regolith.

\subsection{Core model results}

\begin{figure}
\begin{center}

\includegraphics[width=.6\textwidth]{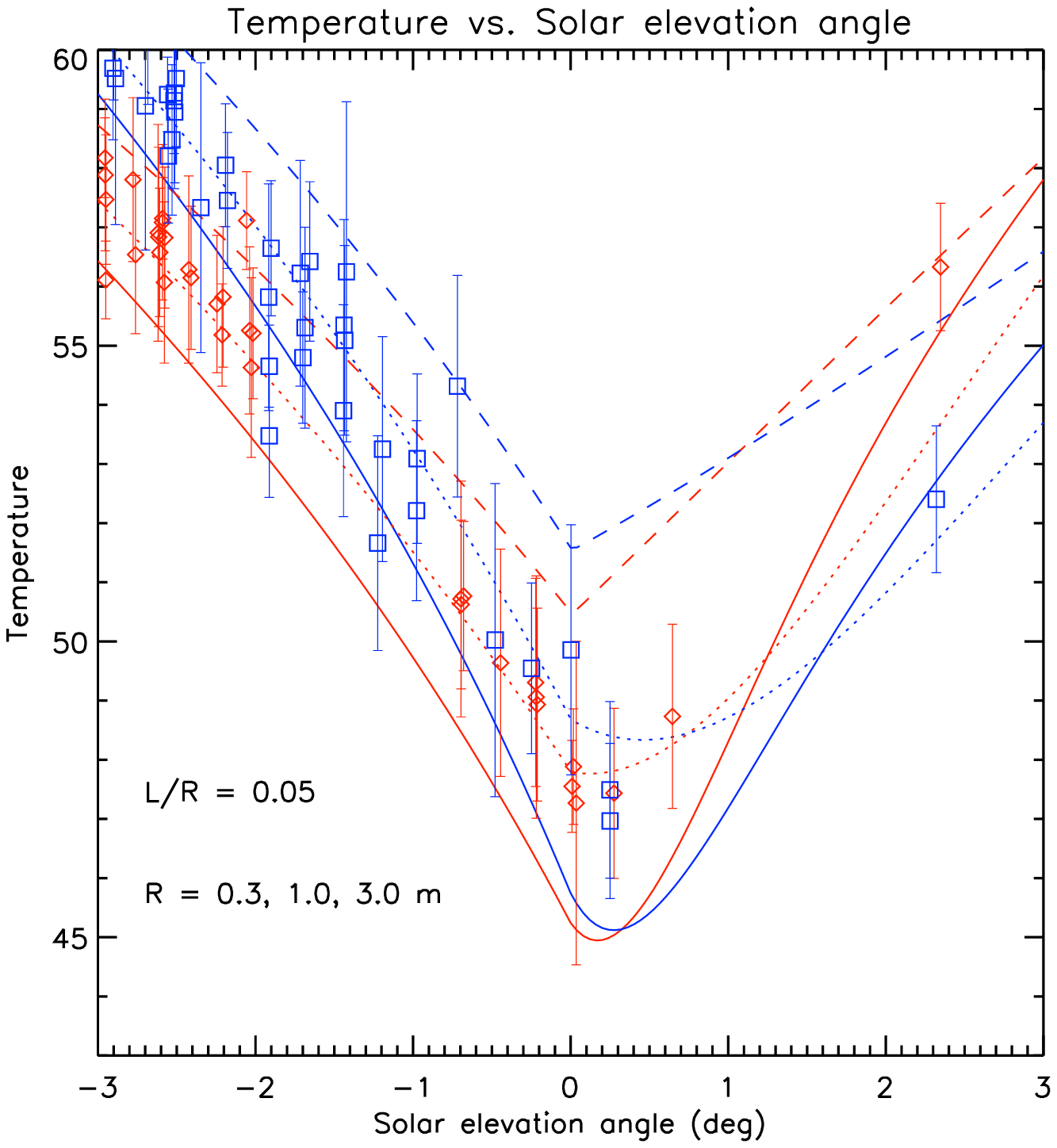}
\includegraphics[width=.6\textwidth]{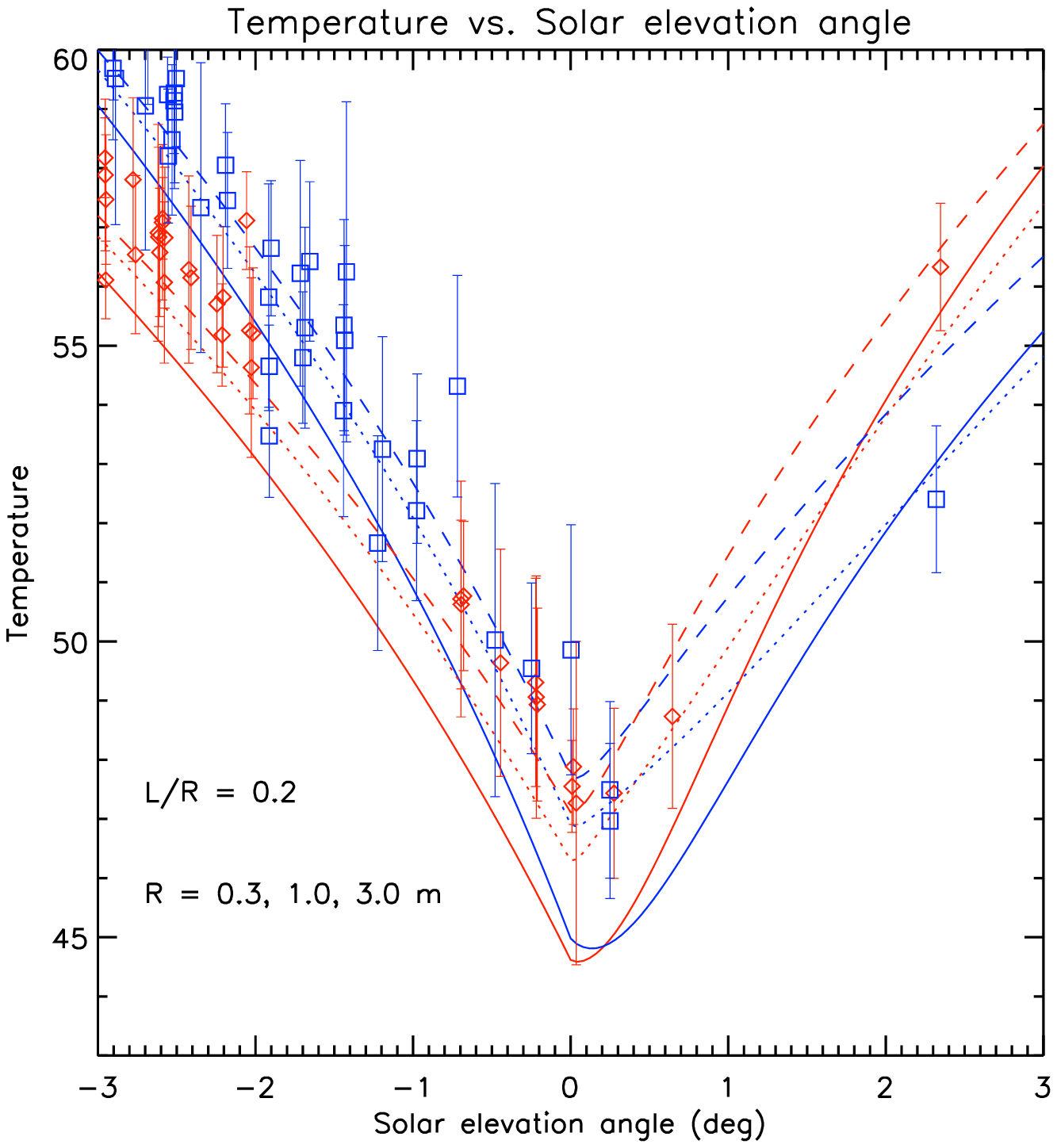}

\end{center}

Fig.~15. 
Same as Fig.~10 but for the core model. 
In the upper and lower panels,  $L/R = 0.2$ and 0.05, respectively.
The results of three different radii $R = 0.3$(solid), 1.0(dotted), and 3.0 m (dashed) are shown 
for the southern (blue) and northern (red) faces.

\end{figure}

\begin{figure}
\begin{center}

\includegraphics[width=.8\textwidth]{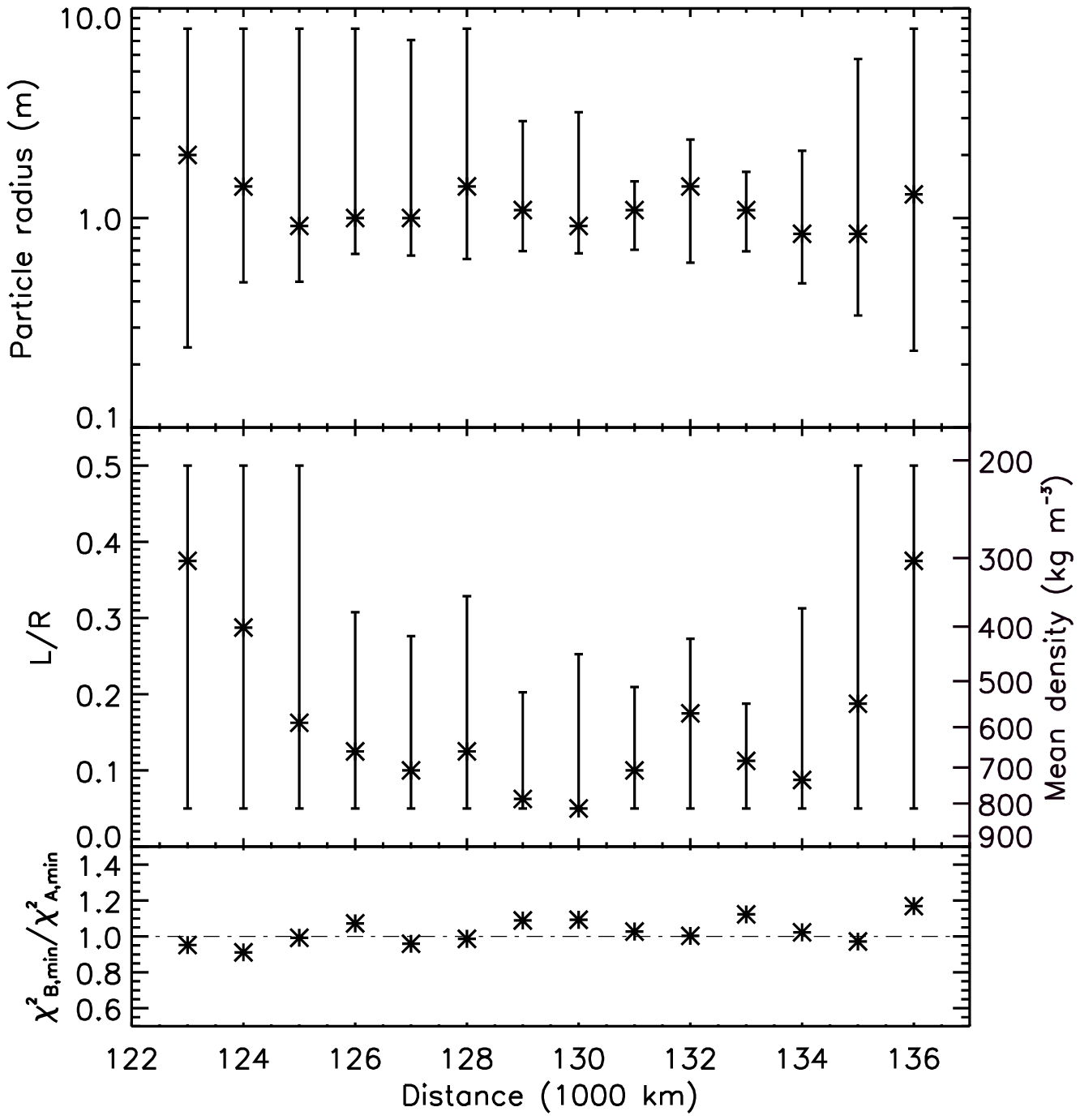}

\end{center}

Fig.~16. 
Best-fit parameters for the core model. Top: particle radius $R$.  Middle: ratio of regolith thickness to particle radius, $L/R$.
Bottom: goodness of fits, $\chi_{\rm B, min}^2/\chi_{\rm A, min}^2$, relative to that of the uniform model.
In the middle panel, the mean internal density is also shown on the right axis.
The ranges of $R$ and $L/R$ are limited a priori in data fits as 0.125 m $\le R \le$ 8 m and 0.05 $\le L/R \le $ 0.5.

\end{figure}

\begin{figure}
\begin{center}

\includegraphics[width=.6\textwidth]{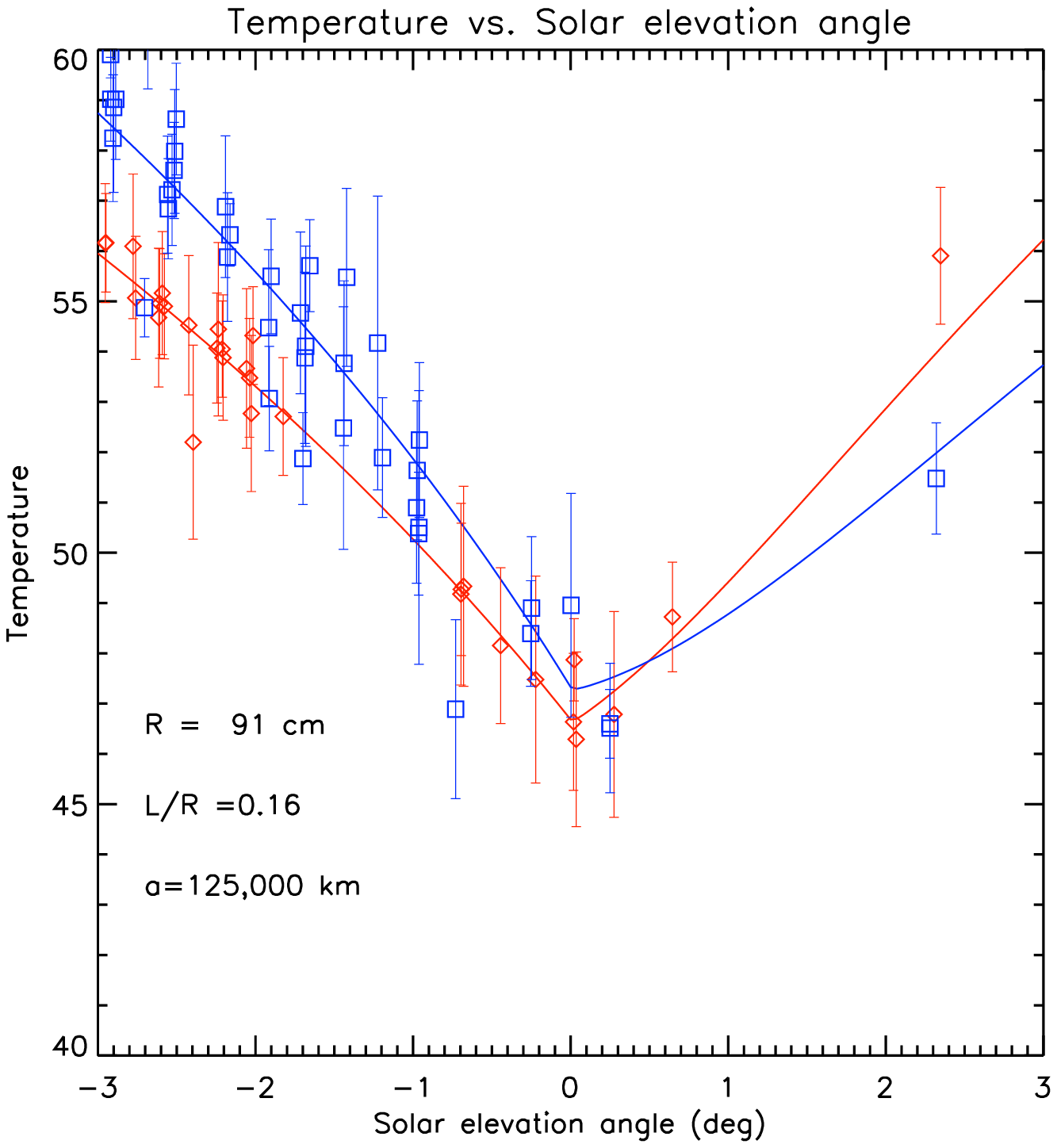}
\includegraphics[width=.6\textwidth]{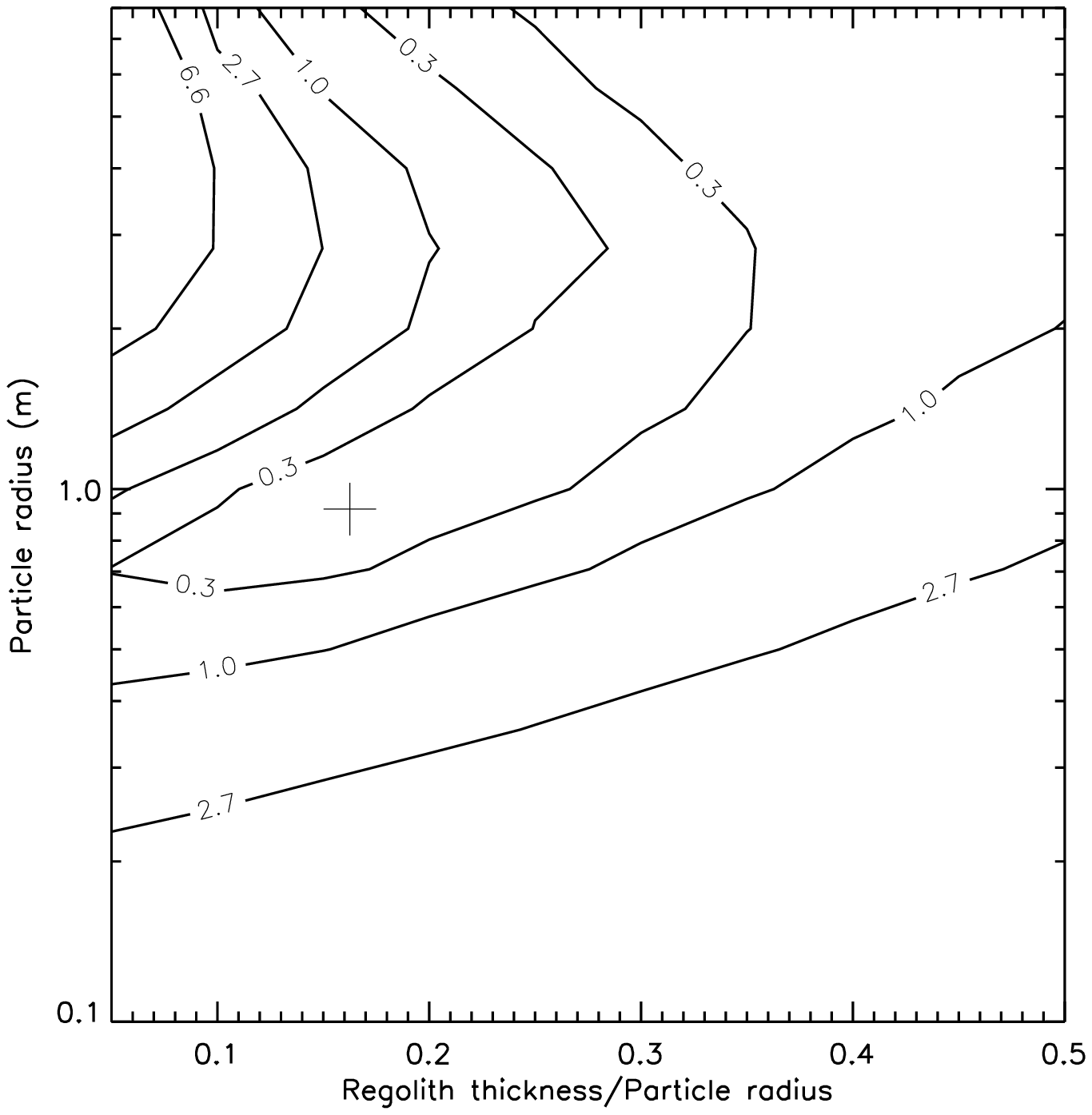}

\end{center}

Fig.~17. 
Data fitting with the core model  at $125,000$ km. 
Top: best-fit temperature curves as a function of $B'$.
The northern and southern face temperatures are represented by red and blue colors, respectively.
Bottom: contour of $\chi_{\rm B}^2/\chi_{\rm B,min}^2-1$ on the plane of $L/R$ vs. $R$. 
The point of the best-fit $L/R$ and $R$ is shown by the plus mark.

\end{figure}

\begin{figure}
\begin{center}

\includegraphics[width=.6\textwidth]{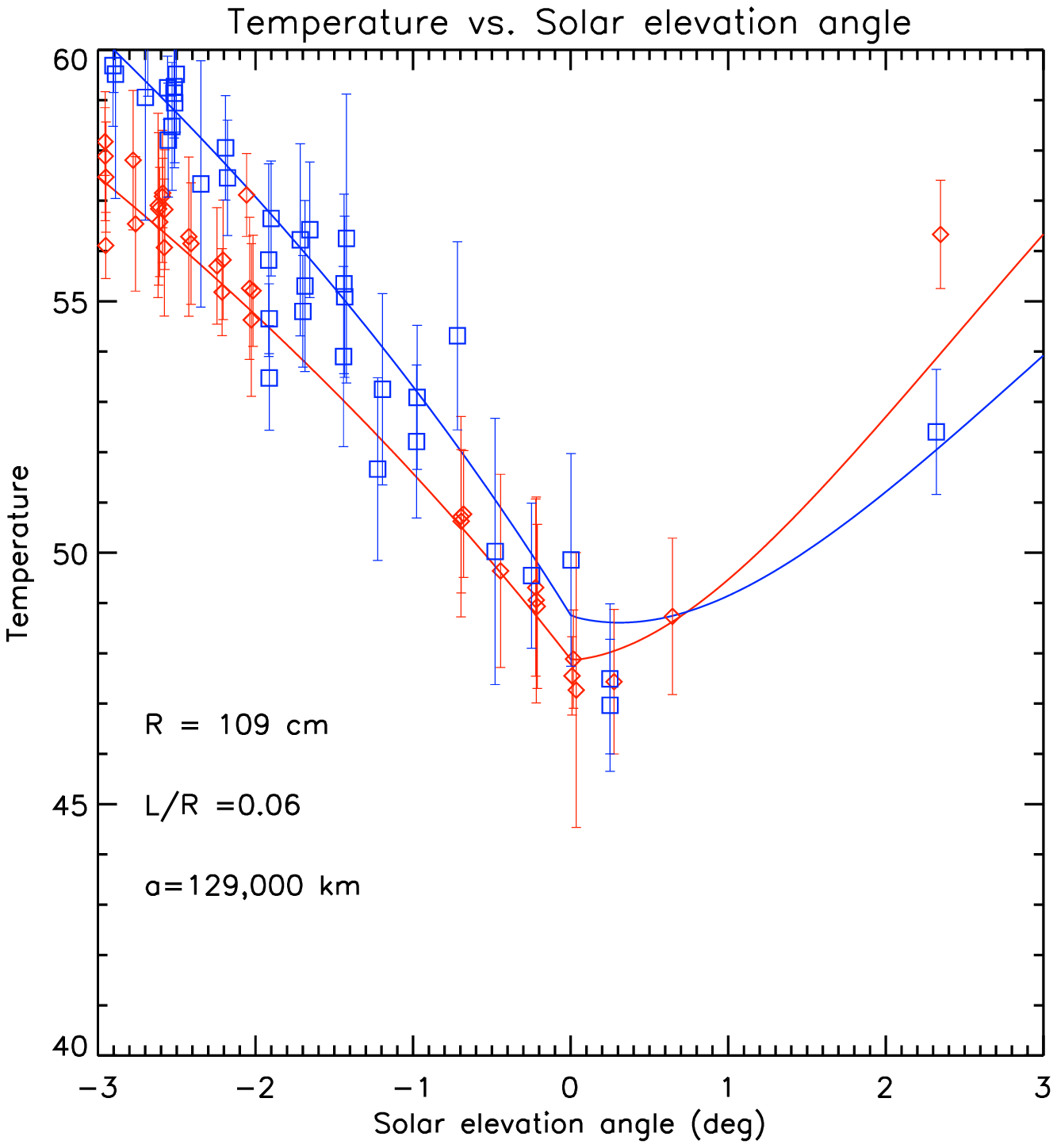}
\includegraphics[width=.6\textwidth]{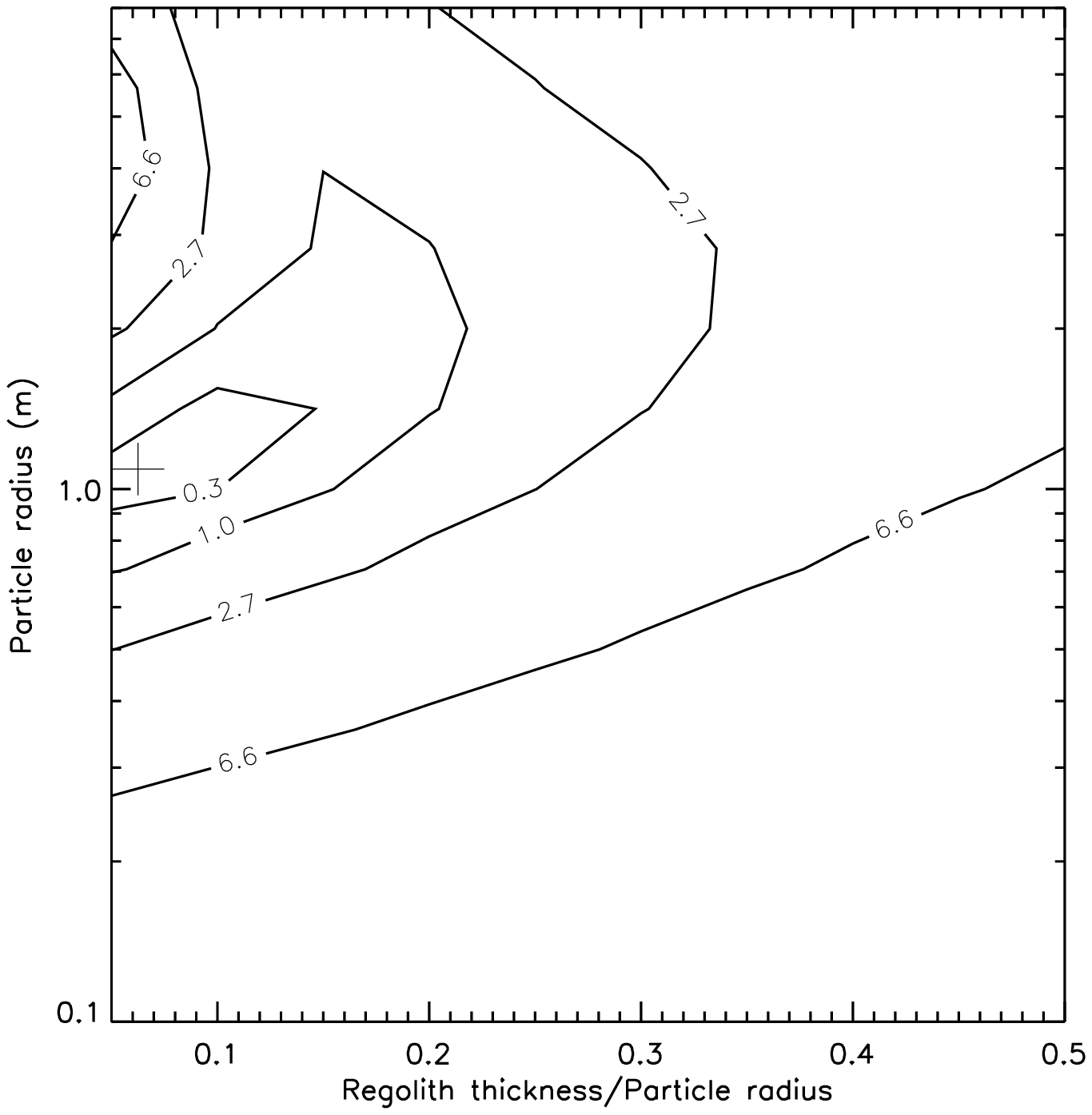}

\end{center}

Fig.~18. 
 Same as Fig.~17, but for the case of $a = 129,000$ km
 
\end{figure}

We first examine basic trends of seasonal temperature variations with different input parameters, as done for the uniform model. 
Figure~15 shows the seasonal temperature curves at $a = 129,000$ km for $L/R = 0.05$ and 0.2.
Three different particle radii,  $R = 0.3$, 1, and 3 m, are considered. The results may be compared with those in Fig.~10 for the uniform model. 
If the regolith thickness is large enough, the temperature curve 
is similar to that for the uniform model with $\Gamma = 10$ ${\rm J \hspace{0.1em} m^{-2} K^{-1} s^{-1/2}}$. 
The case of $L/R = 0.2$ (upper panel) is close to this limit, although the equinox temperatures are slightly increased due to 
the effect of the core.  Since the core cools down slowly and is much warmer than the regolith mantle at the equinox, 
the core warms up the regolith mantle.
With decreasing $L/R$, the equinox temperature increases. This effect is similar to increasing $\Gamma$ in the uniform model.
However,  the shapes of the temperature curves are slightly different between the core model with $L/R = 0.05$ and 
the uniform model with $\Gamma = 50$ ${\rm J \hspace{0.1em} m^{-2} K^{-1} s^{-1/2}}$, even though the modeled equinox temperatures 
are similar to each other.
In the core model, we see a sharp kink at the equinox, particularly for $R = 3$ m, 
as the low $\Gamma$ regolith responds quickly to the incoming flux.
On the other hand,  the temperature minimum comes slightly after the equinox and the temperature curve there is round, if $\Gamma$ is large in the uniform model.

The best-fit parameters for the core model  are shown in Fig.~16.
The particle radius is about 1 m across throughout the A ring, and is slightly smaller than those estimated in the uniform model on average.  
Unlike the uniform model, the core model gives upper limits of $R$ in the middle A ring, 
where relatively large values of $\Gamma$ are seen in the uniform model. 
The regolith thickness $L$ relative to $R$ takes small values in the middle A ring 
whereas it is high in the inner and outer A ring.
In principle,  a small $L/R$ is seen at the location with a high $\Gamma$ in the uniform model.
The mean $\rho$ of the particle calculated from $L/R$ is shown on the right axis of the middle panel. 
The mean $\rho$ is a maximum $\sim 800$ kg m$^{-3}$ in the middle A ring and much lower values of $\rho$ ($\le$ 300 kg m$^{-3}$)
are seen near the edges of the A ring. The value of the mean $\rho$ in the middle A ring is probably 
too high to prevent ring particles from gravitational 
accretion, and a finite porosity seems necessary for the cores.  
The bottom panel of Fig.~16 shows $\chi_{\rm B,min}$ relative to $\chi_{\rm A,min}$. This is about unity at all radial locations,
indicating the model fits are good also for the core model.
  
Examples of best-fit curves and the contours of $\chi_{\rm B}^2/\chi_{\rm B,min}^2-1$ on the $L/R$ vs. $R$ plane are shown in Fig.~17 ($a = 125,000$ km)
and Fig.~18 ($a = 129,000$ km).
At $a = 125,000$ km, the best-fit parameters are $L/R = 0.16$ and $R = 91$ cm.
Although the error bars of individual parameters are large,
we can exclude a combination of a large $R$ and a low $L/R$; 
this case gives an equinox temperature that is too large compared with the observed temperature.
The lower limit of $R$ is well constrained as a small $R$ gives a low equinox temperature.
At $a = 129,000$ km, the best-fit parameters are $L/R = 0.06$ and $R = 109$ cm. 
The contour of Fig.~18 is similar to that in Fig.~17, but a band of low $\chi_{\rm B}^2$ is relatively shifted toward the upper-left corner.
The upper limits of $R$ and $L/R$ are well constrained in this case, as shown in Fig.~16.
It is found that if $R$ is large,  the core continuously warms up 
the regolith mantle even long after the equinox. As a result, regardless of $L/R$, the modeled temperatures, particularly for the south face, 
become higher than the observed temperatures at $B' \sim 2.5^{\circ}$ (see the curves of $R = 3$ m in Fig.~15). 
If $R$ is small ($<$ 1m) this effect is weak and the lower post-equinox temperatures are obtained.

\section{Discussion}

\subsection{Ring particle size}
\begin{figure}
\begin{center}

\includegraphics[width=.65\textwidth]{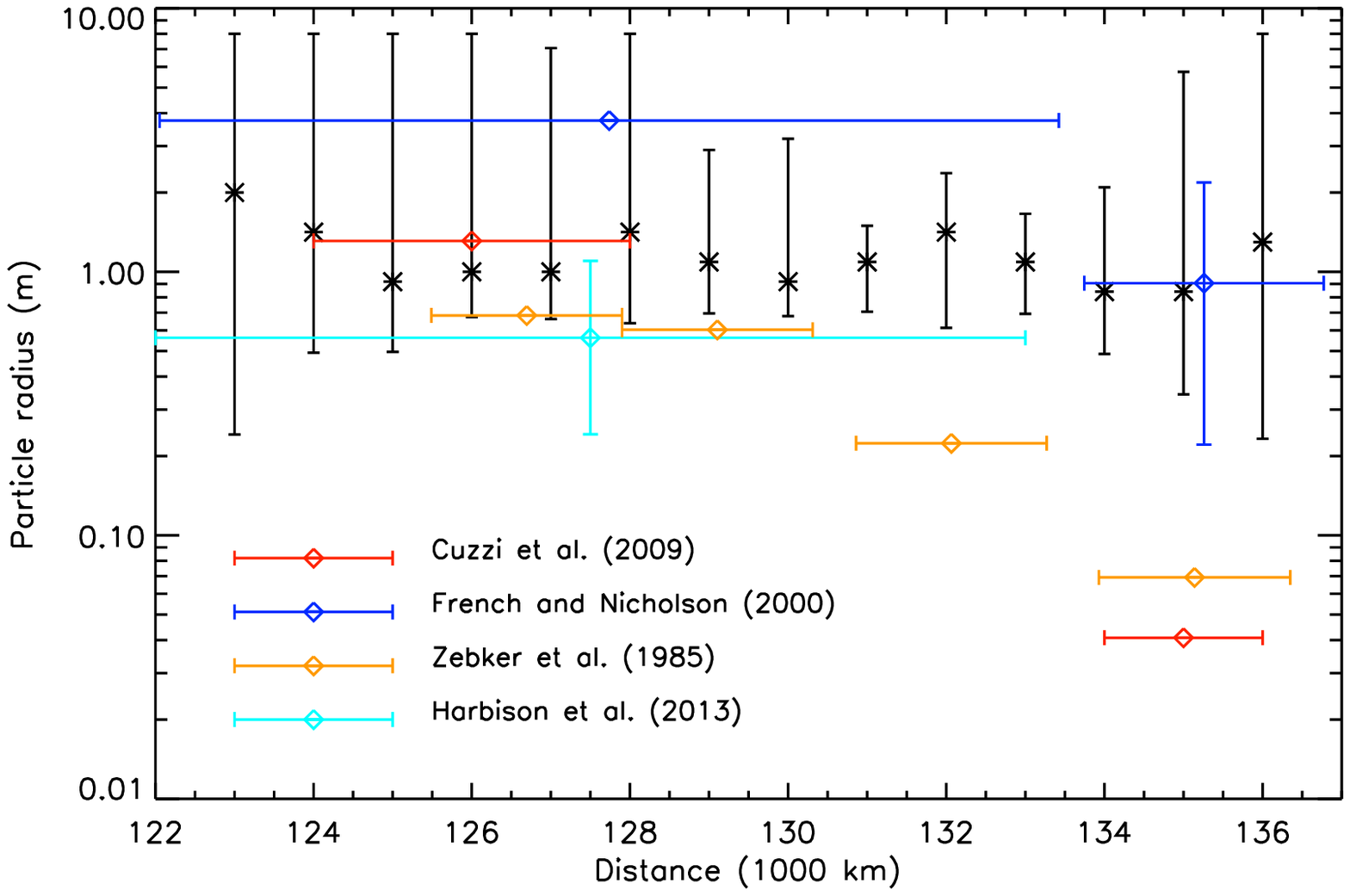}

\end{center}

Fig.~19. Median particle sizes deduced from stellar and radio occultations 
(colored diamonds). The horizontal error bar is the range of saturnocentric distance used for 
averaging of observational data or fitted parameters.
For Cuzzi et al. (2009), we use $R_{\rm min} = 3$ mm, $R_{\rm max} = 10$ m, and $q = 2.7$ for the inner A ring
and $R_{\rm min} = 4$ mm, $R_{\rm max} = 3$ m, and $q = 3.2$ for the outer A ring (their Fig.~15.3).
For French and Nicholson (2000), $q = 2.9$ is used for the outer A ring, while the error bar for 
$R_{\rm med}$ comes from a possible uncertainty in $q$ ($2.75 \le q \le 3.1$), as discussed by them.
For Harbison et al. (2013), $q = 2.8$ is used, while the  the error bar for 
$R_{\rm med}$ comes from a possible uncertainty in $q$ (we adopt $2.7 \le q \le 2.9$); 
$R_{\rm min}$ is altered with $q$ using their Fig.~16.
 The black asterisks are the particle sizes estimated from our core model (Fig.~16).
 
 \end{figure}

The best-fit particle size is about 1 m in our models, although relatively smaller sizes are possible near 
the inner and outer edges of the A ring. 
The particle size has been estimated from various occultations 
(Zebker et al., 1985;  French and Nicholson, 2000;  Cuzzi et al., 2009; Harbison et al. 2013).
These studies showed that  ring particles have a size distribution with an differential power-law exponent of $\simeq -3$,
while our model adopts a single representative size. 
Since the thermal flux is proportional to the total cross section of particles in the absence of mutual shading, 
it may be meaningful to use the median radius of particles $R_{\rm med}$ for comparison, 
where $R_{\rm med}$ is defined so that the total cross section of particles with $R > R_{\rm med}$
is equal to that of  $R < R_{\rm med}$.
The median radius is given as $ [0.5(R_{\rm max}^{3-q}+R_{\rm min}^{3-q})]^{1/(3-q)}$,
where $R_{\rm max}$ and $R_{\rm min}$ are the upper and lower cut off radii, 
and $-q$ is the power-law exponent for the differential size distribution.
The values of $R_{\rm med}$ from occultations are shown in Fig.~19 
together with the particle sizes estimated by our core model (Fig.~16).

For the region inside the Encke division, 
$R_{\rm med}$ from all studies are compatible with our estimates.
For the outer A ring, $R_{\rm med}$ from French and Nicholson (2000) is  
compatible with our estimates whereas 
 $R_{\rm med}$ from Cuzzi et al. (2009) and Zebker et al. (1985) is  much smaller than the lower limit of $R$ 
estimated by our models.
The difference in $R_{\rm med}$ between these studies of occultations comes from
the population of particles smaller than 1 cm; 
$R_{\rm min}$ = 1 cm in French and Nicholson (2000) while $R_{\rm min}$ = 4 mm and 1mm in 
Cuzzi et al. (2009) and Zebker et al. (1985), respectively. 

If the cross section of the outer A ring is dominated by mm to cm-sized particles, as suggested by 
Cuzzi et al. (2009) and Zebker et al. (1985),
the ring should cool down close to the equilibrium temperature at the equinox. 
On the other hand, the CIRS data show the ring temperatures slightly higher than the equilibrium equinox temperatures in the outer A ring (by $\sim$ 3 K). 
This may indicate that our radiative transfer models underestimate the equilibrium equinox temperatures, which we do not believe very likely (Section~3).
The apparent contradiction may be explained if we consider 
possible gravitational aggregates (e.g., Salo, 1995) whose total cross section is small but total mass dominates in the outer A ring.
They can be much warmer than the equilibrium temperature at the equinox. 
If they break up due to mutual collisions (Karjalainen, 2007) for example, small warm particles are dispersed. 
If some of aggregates are persistent on a time scale longer than the thermal relaxation time ($\sim$ a year),
the warm equinox temperature may be explained.

\subsection{Radial variation of internal density}
The temperature anomaly at the equinox is prominent in the middle A ring 
and this was interpreted by a relatively large thermal inertia for the uniform model and 
a relatively large core size in the core model. As pointed out in Section~4.5.1, 
the uniform model is not self-consistent as a very small $K$ estimated by the model 
is not expected for a medium with a moderately high density suggested from self-gravity wakes.
Therefore, we prefer the interpretation of the core model to that of the uniform model.
Namely, the internal density of ring particles, $\rho$, is highest in the middle A ring and it decreases 
towards the inner and outer edges of the A ring.

Some other observations seem to support this idea.
The amplitude of azimuthal brightness asymmetry in photometric observations has a peak in the middle A ring (Dones et al., 1993; French et al., 2007) 
and it sharply decreases towards the inner and outer edges of the A ring.
French et al. (2007) suggested several possibilities to produce the relatively weak asymmetric feature in the inner A ring 
as compared with the middle A ring:
$\rho$ is lower, the particle size distribution is broader, or the restitution coefficient 
is higher in the inner A ring. Among them,  any clear difference in the size distributions between the inner and middle regions of the A ring 
is not identified (Zebker et al., 1985; French and Nicholson, 2000). 
Also, similar elastic properties of ring particles are expected in these regions
because  similar surface regolith properties are suggested from ring spectra (e.g., Hedman et al., 2013).
Therefore, the radial $\rho$-variation seems to be the most plausible cause of 
the steep radial change of the azimuthal brightness asymmetry.
Low-$\rho$ particles in the inner A ring are also indicated from 
existence of axisymmetric wakes (Thomson et al., 2007; Hedman et al. 2014), which 
are probably produced by overstability that is induced if $\rho$ is low (Salo et al., 2001).
It is unclear whether low-$\rho$ particles are necessary for the weak 
azimuthal brightness asymmetry seen in the outer A ring because
other explanations may work, such as a broad size distribution,
particle clumping rather than wake formation, and velocity enhancement due to satellite perturbations.

Cassini ISS observations identified many propeller-forming moonlets  in the A ring 
(Tiscareno et al., 2006, 2008, 2010; Srem\v{c}evi\'{c} et al., 2007). 
They are particularly clustered in the middle A ring  (Srem\v{c}evi\'{c} et al., 2007; Tiscareno et al., 2008) 
that is around the equinox temperature peak.
These moonlets probably have dense cores surrounded by relatively less dense particles as suggested for 
small inner satellites (Porco et al., 2007; Yasui et al., 2014).
It is natural that dense particles are also relatively abundant in the background of the middle A ring.  

The question is what causes the radial variation of $\rho$ or regolith thickness, if it is real. 
Unless some mechanisms work, the radial variation of $\rho$ must be smoothed out
on the viscous diffusion time, $\sim 10^{8}$ years for the A ring (Esposito, 1983; Tiscareno et al., 2007).
Some explanations can be considered for the difference between the middle and outer regions of the A ring.
Relatively abundant small particles in the trans-Encke region (French and Nicholson, 2000, Cuzzi et al., 2009)
indicate that mutual impact velocities are large there probably due to strong self-gravity wakes or satellite perturbations.
It is likely that the dust production rate in such an environment is high and 
ring particles are covered by thick regolith. 
Accumulation of E ring particles may also be important near the outer edge of the A ring (Hor\'{a}nyi et al., 2009). 

The more puzzling mystery is the difference between the inner and middle regions of the A ring. 
We speculate that moonlets selectively transport high-$\rho$ particles outward.
Moonlets are known to radially migrate due to collisional and gravitational interactions with 
surrounding particles, although the exact migration mechanism is under debate
(Crida et al. 2010; Pan and Chiang, 2010; Rein and Papaloizou, 2010; Tiscareno, 2012; Pan et al. 2012; Bromley and Kenyon, 2013).
We do not discuss the detailed mechanisms of migration, but simply consider a situation that 
moonlets can migrate a long distance 
either inward or outward on a long timescale due to gravitational torques. 
If moonlets are rubble-pile aggregates, however, 
inward migration may be inefficient as they break up due to the tidal force. 
Such a restriction does not exist for outward migration, although 
moonlets may not be able to migrate beyond some strong satellite resonances (Bromley and Kenyon, 2013)
and are trapped in certain regions which may be currently seen as the moonlet belts in the middle A ring. 
After moonlets migrate outward,  some of them 
may break up due to mutual collisions or meteoroid impacts and disperse high-$\rho$ particles.

An alternative interpretation for the radial $\rho$ variation 
is that the A ring, or at least the middle part of it, is simply younger than the viscous diffusion time.
For example, a large icy satellite ($\sim$ 100 km in size) 
might have existed at  the present location of the middle A ring and been destroyed by a meteoroid impact. 
The dispersed chunks of dense material have not had enough time to diffuse throughout the A ring.
A similar discussion is found for radial diffusion of moonlets (Crida et al., 2010).
A young A ring is also indicated from other various estimates (Esposito, 1986; Charnoz et al., 2009 and references therein).

In addition to them,  thin regolith mantles found in the present work provide a new constraint on the A ring age. 
Elliott and Esposito (2011) show that the regolith depth of ring particles increases 
with time due to meteoritic bombardment if ring particles are originally regolith-free. 
If we apply the regolith thickness of 10 cm for 1 m particles (Fig.~16) to 
the time evolution of the regolith thickness (their Fig.~2),
the ring age is found to be only $\sim 10^5$ yr.
Since the meteoroid bombardment flux is recently found to be much smaller (Kempf et al., 2013)
than the previous estimate (Cuzzi and Estrada, 1998), the timescale may be larger by 
one to two orders of magnitude.  This is still much shorter than the age of the solar system and is close to that 
suggested from the viscous diffusion time.  

A problem of the young A ring hypothesis 
is that the size of an impactor that can destroy
a 100 km body is $\sim 10$ km and that 
the flux of such massive bodies is very low;
the probability of destruction of a 100 km body in 10$^8$ yr is only $\sim 10^{-3}$ 
(Charnoz et al., 2009). 
In this estimation, the critical impact energy necessary to disperse a half target mass in free space is used
(Benz and Asphaug, 1999). Impact simulations including the tidal force of the planet
(Karjalainen, 2007; Hyodo and Ohtuski, 2014), however,   
show that impact outcomes at the A ring location are much more erosive than those in free space, if the target is a gravitational aggregate.
Thus,  destruction of a large satellite may not be an unrealistically low-probability event.

%(variation in trance Encke exists)

\subsection{Composition of particle cores}
We assumed rigid water ice for the core material in the core model. 
However, similar values of $R$ and $L/R$ can be derived even for other materials, 
as long as $\rho C$ for the core is similar to the assumed value [see Eq.~(\ref{eq:core})].
In fact,  the values of $\rho C$ of chondrites (or rocks) are similar to that for water ice;
as compared with water ice, their densities are larger by a factor of  3-4 but the specific heats are lower by a factor of 3-4 
 (Yomogida and Matsui, 1983; Consolmagno et al., 2013).
If the cores of ring particles are chondritic and $L/R \sim 0.9$, the mean density 
results in $\sim$ 3,000 kg m$^{-3}$. 
Such high density particles cannot exist as ring particles at the A ring location, because accretion must quickly occur.
Therefore, we can exclude cores predominantly made of rocks.
The cores must be predominantly water ice, probably with finite porosities.

\subsection{Other rings}
As found in Spilker et al. (2013), the equinox temperatures for the B and C rings are 
well reproduced by the models with the Saturn flux only.  For the B ring, 
we also check the asymmetry between the pre- and post-equinox temperatures, 
as done for the A ring (Figs.~7 and 8).
We find that the pre-equinox temperatures are higher than the post-equinox temperatures  
for the unlit face, but it is opposite for the lit face, unlike the middle A ring. 
These facts indicate that
the equinox temperature anomaly for the B ring is much smaller than that for the middle A ring, 
even if it exists. This means that the B ring particles are either smaller or less dense than 
the middle A ring particles. 
Since the particle sizes for the B ring estimated from stellar and radio occultations (French and Nicholson, 2000; Cuzzi et al., 2009)
are almost the same with  A ring particles inside the Encke division, it is likely that  the internal density of B ring particles is 
lower than that for middle A ring particles or may be similar to inner A ring particles.
It is difficult to say anything about the temperature asymmetry for the C ring
because the equinox temperatures strongly depend on observational geometry. 
The correction of the geometry dependence remains for future work.

\section{Conclusion}
In the present paper, we examined the seasonal temperature variation of Saturn's A ring.
The lowest temperature was observed at the solar equinox in August 2009. 
We used two different radiative transfer models (a multilayer model and a mono-layer wake model)
and found that both models give the equinox temperatures that are much lower than 
the observed equinox temperatures, as long as only the flux from Saturn is taken into account. 
We also found that the post-equinox temperatures 
are lower than the pre-equinox temperatures at the same absolute solar elevation angle. 
These facts strongly indicate that the A ring was not completely cooled down at the equinox due to the effect of a finite seasonal 
thermal inertia.

Since the seasonal thermal skin depth is comparable to the ring particle size, 
it is possible to give constraints on the physical properties of particles down to 
the deep interior using the seasonal temperature variation curves.
We developed a simple model for seasonal temperature variation, 
and time dependence was taken into account by solving the thermal diffusion equation for the temperature evolution of the particle interior.  
The model calculates only the lit and unlit face temperatures, ignoring geometry dependence and diurnal variation.
We considered two types of the internal structure of ring particles.
In the first model, uniform internal density and thermal inertia of a ring particle were assumed. 
The particle size was estimated to be 1-2 m. 
The seasonal thermal inertia was found to be $\sim$ 30-50 Jm$^{-2}$K$^{-1}$s$^{-1/2}$ in the middle A ring whereas it 
is $\sim$ 10 Jm$^{-2}$K$^{-1}$s$^{-1/2}$ or as low as  the diurnal thermal inertia in the inner and outermost regions of the A ring. 
In the second model, 
a ring particle has a high density core surrounded by a much less dense regolith mantle.
This model showed that the particle size is still about 1 m and that
the core radius relative to the particle radius is about 0.9 for the middle A ring and is much less for the inner and outer A ring.  
This means that a radial variation of the internal density of ring particles exists across the A ring. 

To confine high density particles in the middle A ring against viscous diffusion, some mechanisms 
may be working. We speculated that moonlets are transporting dense particles to the limited regions.
An alternative idea is that the A ring was recently formed ($<$ 10$^{8}$ yr) by 
catastrophic destruction of a pre-existing icy satellite, so that dense particles have not yet diffused over 
the A ring and regolith mantles of particles have not grown thick. 
We also excluded cores predominantly made of rocks, as such particles have very large internal densities.

\section*{Acknowlegements}
We thank anonymous reviewers for helpful comments that significantly improved the manuscript. 
This research was carried out at the Jet Propulsion Laboratory, California Institute of Technology, 
under contract with NASA. Government sponsorship acknowledged. 
This work was also supported in part by the NASA's OPR Program (NNX14AO36G). 
The numerical simulations were performed using the JPL supercomputer, Aurora. 

\section*{Appendix~A}

In the seasonal model introduced in Section~4,  
the heliocentric distance $r_{\rm S}$ and the solar elevation angle $B'$ as a function of time, $t$, are necessary.
We simply assume that Saturn has a Keplerian orbit around the Sun and 
adopt the orbital elements at the epoch J2000 \footnote{http://nssdc.gsfc.nasa.gov/planetary/factsheet/saturnfact.html};
 the semimajor axis of Saturn $a_{\rm S}$ is 9.537 AU and the orbital eccentricity $e_{\rm S}$ is 0.054.
 The heliocentric distance
 $r_{\rm S}$ is given as (Murray and Dermott, 1999)
 \begin{equation}
 r_{\rm S} = a_{\rm S}(1-e_{\rm S}\cos{E_{\rm S}}),
 \end{equation}  
 where $E_{\rm S}$ is the eccentric anomaly of Saturn.
 The eccentric anomaly $E_{\rm S}$ as a function of time is 
 given by solving the Kepler's equation:
 \begin{equation}
\omega_{\rm S}(t-t_{\rm ph}) = E_{\rm S} - e_{\rm S}\sin{E_{\rm S}}, 
 \end{equation}
where  $\omega_{\rm S} = 2\pi/T_{\rm S}$ is the mean motion of Saturn 
(where $T_{\rm S} =$ 10759.22 day is the orbital period of Saturn)
 and $t_{\rm ph}$ is the time at the perihelion passage. 
 Since $r_{\rm S} = $9.433 AU and Saturn was receding from the Sun at the equinox, 
$E_{\rm S}$ at the equinox was 1.367.
We define $\omega_{\rm S}t  = \pi$ at the equinox, and that gives $\omega_{\rm S}t_{\rm ph}$ =  1.827.
The angle between the vector pointing towards Saturn from the Sun and the Saturn spin vector is 
defined to be $\gamma$. Using $\gamma$, the solar elevation angle $B'$ is defined as 
$B' = \gamma -\pi/2$ and  given as 
\begin{equation}
\sin{B'} = \sin{B'_{\rm max}}\sin{(f_{\rm S}-f_{\rm S,1})} \hspace{0.3em},
\end{equation}
where $B'_{\rm max} = 26.73^{\circ}$ is the obliquity of Saturn or the maximum solar elevation angle.
In the above, $f_{\rm S}$ is the true anomaly of Saturn given as 
\begin{equation}
\cos{f_{\rm S}} = \frac{\cos{E_{\rm S}}-e_{\rm S}}{1-e_{\rm S}\cos{E_{\rm S}}},
\end{equation}
and $f_{\rm S,1} = 1.421$ is  the value at the equinox. Note that $B'$ increases 
during the half orbit of Saturn in which Cassini observations have been made.

%\begin{center}

\section*{REFERENCES}
%\end{center}

\begin{description}
%\item 
%Altobelli, N., Spilker, L.J., Pilorz, S., Brooks, S., Edgington, S., Wallis, B., Flasar, M., 2007.
%C ring fine structures revealed in the thermal infrared.
%Icarus 191, 691--701. 

\item 
Altobelli, N., Spilker, L.J., Leyrat, C., Pilorz, S.,  2008.
Thermal observations of Saturn's main rings by Cassini CIRS: 
Phase, emission and solar elevation dependence.
Planet. Space. Sci. 56, 134--146.

\item 
Altobelli, N., Spilker, L.J., Pilorz, S., Leyrat, C., Edgington, S., Wallis, B., Flandes, A., 2009.
Thermal phase curves observed in Saturn's main rings by Cassini-CIRS: Detection of an opposition effect?
Geophys. Res. Lett. 36, L10105.

\item
Altobelli, N., et al., 2014.
Two numerical models designed to reproduce Saturn ring temperatures as measured by 
Cassini-CIRS.
Icarus 238, 205--220.

\item
Andersson, O., Inaba, A., 2005.
Thermal conductivity of crystalline and amorphous ices and its implications
on amorphization and glassy water.
Phys. Chem. Chem. Phys., 2005, 7, 1441-1449. 

\item
Benz, W., Asphaug, E., 1999.
Catastrophic disruptions revisited.
Icarus 142, 5--20.

\item 
Bromley, C.B., Kenyon, S.J., 2013.
Migration of small moons in Saturn's rings.
Astro. Phys. J., 764, 192.

\item
Chandrasekhar, S., 1960.
Radiative transfer.
Dover, New york. 

\item
Charnoz, S., Dones, L., Esposito, L., Estrada, P.R., Hedman, M.M., 2009.
Origin and evolution of Saturn's ring system.
In: Dougherty, M.K., Esposito, L.W.,  Krimigis, S.M. (Eds.), Saturn from Cassini-Huygens.
Springer, Berlin, pp. 537--575.

\item
Charnoz, S., Salmon, J., Crida, A., 2010.
The recent formation of Saturn's moonlets from viscous spreading of the main rings.
Nature 465, 752--754.

\item
Charnoz, S., et al., 2011.
Accretion of Saturn's mid-sized moons during the viscous spreading of young massive rings:
Solving the paradox of silicate-poor rings versus silicate-rich moons.
Icarus 216, 535--550.

\item 
Colwell, J.E., Esposito, L.W., Srem\v{c}evi\'{c}, M., 2006.
Self-gravity wakes in Saturn's A ring measured by stellar 
occultations from Cassini.
Geophys. Res. Lett. 33, L07201.

%\item 
%Colwell, J.E., Esposito, L.W., Srem\v{c}evi\'{c}, M., Stewart, G.R., McClintock, W.E.,  2007.
%Self-gravity wakes and radial structure of Saturn's B ring.
%Icarus, 190, 127--144.

\item
Colwell, J.E., Esposito, L.W., Jerousek, R.J., Srem\v{c}evi\'{c}, M., Pettis, D., Bradley, E.T., 2010.
Cassini UVIS stellar occultation observations of Saturn's rings.
Astron. J. 140, 1569--1578.

\item
Consolmagno, G.J., Schaefer, M.W., Schaefer, B.E., Britt, D.T., Macke, R.J., Nolan, M.C., Howell, E.S., 2013.
The measurements of meteorite heat capacity at low temperatures
using nitrogen vaporization.
Planet. Space Sci. 87, 146--156.

\item 
Crida, A., Papaloizou, J.C.B., Rein, H., Charnoz, S., Salmon, J., 2010.
Migration of a moonlet in a ring of solid particles: Theory and application to Saturn's propellers.
Astron. J. 140, 944-953.

\item
Cuzzi, J. N.,  Estrada, P.R., 1998.
Compositional evolution of Saturn's rings due to meteoroid bombardment.
Icarus 132,  1--35.

\item
Cuzzi, J.N., Clark, K., Filacchione, G., French, R., Johnson, R., Marouf, E., Spilker, L., 2009.
Ring particle composition and size distribution.
In: Dougherty, M.K., Esposito, L.W.,  Krimigis, S.M. (Eds.), Saturn from Cassini-Huygens.
Springer, Berlin, pp. 459--509.

%\item
%Daisaka, H., Ida, S., 1999. 
%Spatial structure and coherent motion in dense planetary rings 
%induced by self-gravitational instability. Earth Planets Space 51,
%1195--1213.

\item
Daisaka, H., Tanaka, H., Ida, S., 2001.
Viscosity in a dense planetary ring with self-gravitating particles.
Icarus 154, 296--312.

\item 
Deau, E., 2015. 
The opposition effect in Saturn's main rings as seen by Cassini ISS: 
2. Constraints on the ring particles and their regolith with analytical radiative transfer models.
Icarus, in press.

\item 
Dones, L., 1991.
A recent cometary origin for Saturn's rings?
Icarus 92, 194--203.

\item 
Dones, L., Cuzzi, J.N., Showalter, M.R., 1993.
Voyager photometry of Saturn's A ring.
Icarus 105, 184--215.

%\item
%Dunn, D.E., Molnar, L.A., Niehof, J.T., de Pater, I.,  Lissauer, J.L., 2004.
%Microwave observations of SaturnÕs rings: Anisotropy in directly 
%transmitted and scattered saturnian thermal emission. Icarus 171, 183--198.

\item
Dunn, D.E., de Pater, I., Molnar, L.A., 2007.
Examining the wake structure in Saturn's rings from microwave observations over varying
ring opening angles and wavelengths.
Icarus 192, 56--76. 

\item
Elliot, J.P., Esposito, L.W., 2011.
Regolith depth growth on an icy body orbiting Saturn and evolution 
of bidirectional reflectance due to surface composition changes.
Icarus 212, 268--274.

\item
Esposito, L., 1986.
Structure and evolution of Saturn's rings.
Icarus 67, 345--357.

\item 
Esposito, L. et al., 1983.
Voyager photopolarimeter steller occultation of Saturn's rings.
J. Geophys. Res. 88, A11.

\item
Feistel, R., Wagner, W., 2006.
A new equation of state for H$_2$O ice Ih.
J. Phys. Chem. Ref. Data 35, 1021--1047.

%\item 
%Ferrari, C., Leyrat, C., 2006.
%Thermal emission of spherical spinning ring particles: The standard model.
%Astron. Astrophys. 447, 745--760.

%\item 
%Ferrari, C., Reffet, E., 2013.
%The dark side of Saturn's B ring: Seasons as clue to its structure.
%Icarus 223, 28--39.

\item   
Ferrari, C., Galdemard, P., Lagage, P.O., Pantin, E., Quoirin, C., 2005.
Imaging Saturn's rings with CAMIRAS: thermal inertia of B and C rings.
Astron. Astrophys. 441, 379--389.

\item 
Ferrari, C., Brooks, S., Edgington, S., Leyrat, C., Pilorz, S., Spilker, L., 2009.
Structure of self-gravity wakes in Saturn's A ring as measured by Cassini CIRS.
Icarus 199, 145--153.

\item
Flandes, A., Spilker, L.J., Morishima, R., Pilorz, S., Leyrat, C., Altobelli, N., Brooks, S.M., Edgington, S.G., 2010.
Brightness of Saturn's rings with decreasing solar elevation.
Planet and Space Science 58, 1758--1765. 

%\item 
%Flasar, F.M., and 44 colleagues, 2004.
%Exploring the Saturn system in the thermal infrared: The composite infrared spectrometer.
%Space Science Rev. 115, 169--297.

%\item 
%Flasar, F.M., and 45 colleagues, 2005.
%Temperatures, winds, and composition in the Saturnian system.
%Science 307, 1247--1251.

\item 
French, R.G., Nicholson, P.D., 2000.
Saturn's rings II.
Particle sizes inferred from stellar occultation data.
Icarus 145, 502--523.

\item
French, R.G., Salo, H., McGhee, C.A., Dones, L., 2007.
HST observations of azimuthal asymmetry in Saturn's rings.
Icarus 189, 493--522.

\item 
Froidevaux, L. 1981.
Saturn's rings: Infrared brightness variation with solar elevation.
\textit{Icarus} {\bf 46}, 4--17.

\item
Goldreich, P., Tremaine, S., 1982.
The dynamics of planetary rings.
Annu. Rev. Astron. Astrophys. 20, 249--283. 

\item
Harbison, R.A., Nicholson, P.D., Hedmann, M.M., 2013.
The smallest particles in Saturn's A and rings. 
Icarus 126, 1225--1240.

\item 
Hedman, M.M., Nicholson, P.D., Salo, H., Wallis, B.D. Buratti, B.J.,
Baines, K.H., Brown, R.H., Clark, R.N.,  2007.
Self-gravity wake  structures in Saturn's A ring revealed by Cassini VIMS.
Astron. J. 133, 2624--2629.

\item
Hedman, M.M., Nicholson, P.D., Cuzzi, J.N., Clark, R.N., Filacchione, G., Capaccioni, F., Ciarniello, M., 2013.
Connections between spectra and structure in Saturn's main rings based on Cassini VIMS data.
Icarus 223, 105--130.

\item
Hedman, M.M., Nicholson, P.D., Salo, H., 2014.
Exploring overstabilities in Saturn's A ring using two stellar occultations. 
Astron. J. 148, 15.

\item
Hor\'{a}nyi, M., Burns, J.A., Hedman, M.M., Jones, G.H., Kempf, S., 2009.
Diffuse rings.
In: Dougherty, M.K., Esposito, L.W.,  Krimigis, S.M. (Eds.), Saturn from Cassini-Huygens.
Springer, Berlin, pp. 511--536.

\item 
Hyodo, R., Ohtsuki, K., 2014.
Collisional disruption of gravitational aggregates in the tidal environment.
Astrophys. J. 787, 56.

\item
Judson, A., Doesken, N., 2000.
Density of freshly fallen snow in the central rocky mountains.
Bull. Amer. Meteor. Soc. 81, 1577-1587.

\item
Karjalainen, 2007.
Aggregate impacts in Saturn's rings.
Icarus 189, 523-537.

\item
Kempf, S., Altobelli, N., Hor\'{a}nyi, M., Srama, R., 2013.
The mass flux of micrometeoroids into the Saturnian system.
American Geophysical Union, Fall Meeting 2013, abstract P21E-05. 

\item
Krause, M., Blum, J., Skorov, Yu.V., Trieloff, M., 2011.
Thermal conductivity measurements of porous dust aggregates: I. Technique, model and first results.
Icarus 214, 286--296.

\item 
Leyrat, C., Ferrari, C., Charnoz, S., Decriem, J., Spilker, L., Pilorz, S., 2008a.
Spinning particles in Saturn's C ring from mid-infrared
observations: Pre-Cassini results.
Icarus 196, 625--641.

\item
Leyrat, C., Spilker, L, J., Altobelli, N., Pilorz, S., Ferrari, C., 2008b.
Infrared observations of Saturn's rings by Cassini CIRS : Phase angle and local time dependence
Planet. Space. Sci. 56, 117--133.

\item
Li, L.,  et al., 2010.
Saturn's emitted power.
J. Geo. Res.  115, E11002.

%\item
%Lumme, K., Irvine, W.M., 1976. 
%Photometry of SaturnÕs rings. Astron. J. 81, 863--893.

\item
Modest, M., 2003.
Radiative heat transfer.
Academic press.

\item
Morishima, R., Salo, H., Ohtsuki, K., 2009.
A multilayer model for thermal infrared emission of Saturn's rings:
Basic formulation and implications for Earth-based observations.  
Icarus 201, 634--654.

\item
Morishima, R., Spilker, L., Salo, H., Ohtsuki, K., Altobelli, N., Pilorz, S., 2010.
A multilayer model for thermal infrared emission of Saturn's rings. II:
Albedo, spins, and vertical mixing of ring particles inferred from Cassini CIRS.  
Icarus 210, 330--345. 

\item  
Morishima, R., Spilker, L., Ohtsuki, K.,  2011.
A multilayer model for thermal infrared emission of Saturn's rings III:
Thermal inertia inferred from Cassini CIRS.
Icarus 215, 107--127. 

\item  
Morishima, R., Edgington, S.G., Spilker L.,  2012.
Regolith grain sizes of Saturn's rings inferred from Cassini-CIRS far-infrared spectra.
Icarus 221, 888--899. 

\item  
Morishima, R., Spilker L., Turner, N.,  2014.
Azimuthal temperature modulations of Saturn's A ring caused by 
self-gravity wakes.
Icarus 228, 247--259. 

%\item
%Mullin, T.S., 1984. 
%Glossary. In: Gehrels, T., Matthews, M.S. (Eds.), 
%Saturn. Univ. of Arizona Press, Tucson, pp. 931Ð945.

\item
Murray, C.D., Dermott, S.F., 1999. 
Solar System Dynamics. 
Cambridge Cambridge Univ. Press, Cambridge.

%\item
%Nicholson, P.D., Hedman, M.M., 2010.
%Self-gravity wake parameters in Saturn's A and B rings.
%Icarus 206, 410--423.

\item 
Pan, M., Chiang, E., 2010.
The propeller and the frog.
Astro. Phys. J. Lett. 722, L178--L182.

\item 
Pan, M., Rein H., Chiang, E., Evans, S.N., 2012.
Stochastic flight of propellers.
Mon. Not. R. Astron. Soc. 447, 2788--2796.

\item 
Pilorz, S., Altobelli, N., Colwell, J., Showalter, M., 2014.
Thermal transport in Saturn's B ring inferred from Cassini CIRS.
Icarus, in prese,  doi:10.1016/j.icarus.2015.01.002.  

\item
Porco, C.C., Thomas, P.C., Weiss, J.W., Richardson, D.C., 2007.
Saturn's small inner satellites: Clues to their origins.
Science 318, 1602--1607.

\item
Poulet, F., Cruikshank, D.P., Cuzzi, J.N., Roush, T.L., French, R.G., 2003.
Composition of Saturn's rings A, B, and C from high resolution
near-infrared spectroscopic observations.
Astron. Astrophys. 412, 305--316.

\item
Press, W.H., Teukolsky, S.A., Vetterling, W.T., Flannery, B.P., 1986.
Numerical Recipes.
Cambridge Univ. Press, Cambridge, UK.

\item
Rein, H., Papaloizou, J.C.B., 2010.
Stochastic orbital migration of small bodies in Saturn's rings.
Astron. Astrophys. 524, 22.

\item 
Robbins, S.J., Stewart, G.R., Lewis, M.C., Colwell, J.E., Srem\v{c}evi\'{c}, M.,  2010.
Estimating the masses of Saturn's A and B rings from high-optical depth $N$-body simulations and stellar occultations.
Icarus 206, 431--445.

%\item
%Rubincam, D.P., 1995.
%Asteroid orbit evolution due to thermal drag.
%J. Geophys. Res. 100, 1585--1594. 

%\item
%Salo, H., 1992. 
%Numerical simulations of dense collisional systems. II - Extended distribution of particle sizes.
%Icarus 96, 85--106.

\item
Salo, H., 1995. 
Simulations of dense planetary rings. III. Self-gravitating identical particles. 
Icarus 117, 287--312.

%\item
%Salo, H., Karjalainen, R., 2003.
%Photometric modeling of Saturn's rings 
%I. Monte Carlo method and the effect of nonzero volume filling factor.
%Icarus 164, 428--460.

\item
Salo, H., Schmidt, J., Spahn, F., 2001. 
Viscous overstability in Saturn's B-ring. I. 
Direct simulations and measurement of transport coefficients. 
Icarus 153, 295--315.

\item 
Salo, H., Karjalainen, R., French, R.G., 2004.
Photometric modeling of Saturn's rings. II. Azimuthal asymmetry in reflected
and transmitted light.
Icarus 170, 70--90.

\item
Salmon, J., Charnoz, J., Crida, A., Brahic, A., 2010.
Long-term and large-scale viscous evolution of dense planetary rings.
Icarus 209, 771--785. 

\item
Schmidt, J., Tiscareno, M.S., 2013.
Ejecta clouds from meteoroid impacts on Saturn's rings: Constraints on 
the orbital elements and size of projectiles. 
American Geophysical Union, Fall meeting 2013, abstract P23D-1820.

\item 
Srem\v{c}evi\'{c}, M., Schmidt, J., Salo, H., Sei\ss, Spahn, F., Albers, N., 2007.
A belt of moonlets in Saturn's A ring.
Nature 449, 1019--1021. 

\item
Shoshany, Y., Prialnik, D., Podolak, M.,  2002.
Monte Carlo modeling of the thermal conductivity of porous cometary ice.
Icarus 157, 219--227.

%\item
%Spilker, L., Pilorz, S.H., Edgington, S.G., Wallis, B.D.,
%Brooks, S.M., Pearl, J.C., Flasar, F.M., 2005.
%Cassini CIRS observations of a roll-off in Saturn ring spectra at
%submillimeter wavelengths.  
%Earth, Moon, and Planets 96, 149--163. 

\item 
Shulman, L.M., 2004.
The heat capacity of water ice in interstellar or interplanetary conditions.
Aston. Astrophys. 416, 187--190.

\item
Spilker, L. et al., 2006.
Cassini thermal observations of Saturn's main rings: 
Implications for particle rotation and vertical mixing.
Planet. Space Sci. 54, 1167--1176.

\item
Spilker, L., Ferrari, C., Morishima, R., 2013.
Saturn's ring temperatures at equinox.
Icarus 226, 316--322.

\item
Thomson, F.S., Marouf, E.A., Tyler, G.L., French, R.G., Rappoport, N.J., 2007.
Periodic microstructure in Saturn's rings A and B.
Geo. Res. Lett. 34, L24203.

\item
Tiscareno, M.S., Burns, J.A., Hedman, M.M., Porco, C.C., Weiss, J.W., Dones, L., Richardson, D.C., 
Murray, C.D., 2006.
100-metre-diameter moonlets in Saturn's A ring from observations of 'propeller' structures.
Nature 440, 648--650.

\item 
Tiscareno, M.S., Burns, J.A., Nicholson, P.D., Hedman, M.M., Porco, C.C., 2007.
Cassini imaging of Saturn's rings. II. A wavelet technique for analysis of density waves and other 
radial structure in the rings.
Icarus 189, 14--34.

\item 
Tiscareno, M.S., Burns, J.A., Hedman, M.M., Porco, C.C., 2008.
The population of propellers in Saturn's rings.
Astro. Phys. J. 135, 1083--1091.

\item 
Tiscareno, M.S., and 10 colleagues, 2010.
Physical characteristics and non-Keplerian orbital motion of "propeller" moons
embedded in Saturn's rings.
Astro. Phys. J. Lett. 718, L92--L96.  

\item 
Tiscareno, M.S. et al., 2013.
Observations of ejecta clouds produced by impacts onto Saturn's rings.
Science 340, 460--464.

\item
Yasui, Y., Ohtsuki, K., Daisaka, H., 2014.
Gravitational accretion of particles onto moonlets embedded in Saturn's rings.
Astrophys. J., in press.

\item
Yomogida, K., Matsui, T., 1983.
Physical properties of ordinary chondrites.
J. Geophys. Res. 88,  9513--9533.

\item
Zebker, H.A., Marouf, E.A., Tyler, G.L., 1985.
Saturn's rings - Particle size distributions for thin layer model.
Icarus 64, 531--548.

\end{description}

\end{document}